\documentclass[pra,letterpaper,twocolumn,showpacs,superscriptaddress,floatfix]{revtex4-1}
\usepackage{graphicx,psfrag,amsmath,amssymb,amsfonts,bbm,latexsym,color,dcolumn,bm,mathbbol,mathrsfs}
\allowdisplaybreaks

\newcommand{\bra}[1]{\left\langle #1 \right|}
\newcommand{\ket}[1]{\left| #1 \right\rangle}
\newcommand{\I}{\mathrm{i}}
\newcommand{\D}{\mathrm{d}}

\definecolor{myred}{RGB}{168,5,14}
\definecolor{myblue}{RGB}{13,13,255}

\usepackage[normalem]{ulem}

\begin{document}

\title{Thermalization of the Lipkin-Meshkov-Glick model in blackbody
  radiation}

\author{T. Macr\`{i}} \affiliation{Departamento de F\'isica T\'eorica
  e Experimental and International Institute of Physics, Universidade
  Federal do Rio Grande do Norte, Natal-RN, Brazil}
\author{M. Ostilli} \affiliation{Departamento de F\'isica T\'eorica e
  Experimental, Universidade Federal do Rio Grande do Norte, Natal-RN,
  Brazil} \author{C. Presilla} \affiliation{Dipartimento di Fisica,
  Sapienza Universit\`a di Roma, Piazzale A. Moro 2, Roma 00185,
  Italy} \affiliation{Istituto Nazionale di Fisica Nucleare, Sezione
  di Roma 1, Roma 00185, Italy}

\date{\today}

\begin{abstract}
  In a recent work, we have derived simple Lindblad-based equations
  for the thermalization of systems in contact with a thermal
  reservoir.  Here, we apply these equations to the
  Lipkin-Meshkov-Glick model (LMG) in contact with a blackbody
  radiation and analyze the dipole matrix elements involved in the
  thermalization process.  We find that the thermalization can be
  complete only if the density is sufficiently high, while, in the
  limit of low density, the system thermalizes partially, namely,
  within the Hilbert subspaces where the total spin has a fixed value.
  In this regime, and in the isotropic case, we evaluate the
  characteristic thermalization time analytically, and show that it
  diverges with the system size in correspondence of the critical
  points and inside the ferromagnetic region.  Quite interestingly, at
  zero temperature the thermalization time diverges only quadratically
  with the system size, whereas quantum adiabatic algorithms, aimed at
  finding the ground state of same system, imply a cubic divergence of
  the required adiabatic time.
\end{abstract}

\maketitle

\section{Introduction}
The main difference between open and isolated systems, is the lack of
conservation laws in the former, the most common one being the energy
conservation.  For open quantum
systems~\cite{Petruccione,Weiss,Schaller}, another peculiar but less
uniquely defined quantity, quantum coherence, is being loosed.  In
more formal terms, if the system, when isolated, is governed by some
time independent Hamiltonian $\bm{H}$, and if
$\bm{O}_1,\ldots,\bm{O}_q$ are a set of $q$ independent operators that
commute with $\bm{H}$, and commute with each other, the quantum
mechanical averages of these operators, including $\bm{H}$, provide a
set of $q+1$ constants of motion. If instead the system interacts with
some environment, in general, none of these operators is a constant of
motion. Nevertheless, if the system-environment interaction can be
reduced to one of the operators $\bm{O}_1,\ldots,\bm{O}_q$, say
$\bm{O}_1$, then, even if the system looses both energy and quantum
coherence, $\bm{O}_1$ remains conserved during what we can call a
partial thermalization process.  This is what happens in the
Lipkin-Meshkov-Glick (LMG) model~\cite{LMG} when put in contact with a
thermal reservoir constituted by a blackbody radiation at thermal
equilibrium.

It has been proven in~\cite{GK1976} that the reduced density matrix of
a system interacting with a chaotic bath of bosons, which is well
approximated by a blackbody radiation, obeys a Lindblad equation (see
for example~\cite{Petruccione,Weiss,Schaller} and references therein).
Here, by using a Lindblad-based approach (LBA)~\cite{OP,OPprl}, we
analyze the thermalization process of the LMG model embedded in a
blackbody radiation.  The analysis suggests that complete thermal
equilibrium can be reached only at high enough density, while a
partial thermalization takes place at low density. In the latter case,
along the thermalization process, the total angular momenta remains a
conserved quantum number.  We then specialize the analysis of the
thermalization in this low density regime, where the total spin is
conserved.  In the isotropic case, we provide a comprehensive picture
of the characteristic thermalization times, as functions of the
Hamiltonian parameters and of the system size $N$.  Quite importantly,
we find that these characteristic times diverge with $N$ only at the
critical point and in its ferromagnetic phase, linearly at high
temperatures, and quadratically at zero temperature. The latter result
is to be compared with the time estimated for reaching the ground
state of this model by a quantum adiabatic algorithm, which is known
to diverge with $N^3$~\cite{QAD}.

The LMG is a fully-connected model of quantum spins which, in the
thermodynamic limit, is exactly solvable. It has been the subject of
many works, both at equilibrium~\cite{LMG_Botet,LMG_Ribeiro}, along
dynamics afterward a fast quench~\cite{LMG_Das,LMG_Campbell}, along
adiabatic dynamics~\cite{LMG_Santoro}, and in the microcanonical
framework \cite{LMG_Mori}. The LMG model has also been used to
represent an environment of interacting spins in contact with a system
made of a single spin or two spins, by mean-field
approximations~\cite{Paganelli,Paganelli1}, and also by exact
numerical analysis of the reduced density
matrix~\cite{LMG_Petruccione,LMG_Quan}.  The LMG model can find an
approximated experimental realization in certain ferroelectrics,
ferromagnets~\cite{Abragam,Wolf}, and magnetic molecules~\cite{Ziolo}.
In more recent years, the model has attracted a renewed attention due
to the possibility to be simulated by trapped ions~\cite{Cirac}, as
well as by Bose-Einstein condensates of ultracold
atoms~\cite{Cirac98}.  Indeed it has been studied experimentally on
several platforms: with trapped ions \cite{monz11,bohnet16}, with
Bose-Einstein condensates via atom-atom elastic collisions
\cite{gross10,riedel10}, and via off-resonance atom-light interaction
in a optical cavity \cite{leroux10,zhang}.  LMG emerges also as a fully
blockaded limit of Rydberg dressed atoms \cite{henkel10} in lattices
\cite{macri14,jau16,zeiher16}, which could have interesting
applications to quantum metrology \cite{kitagawa93,gil14,macri16} as
well as to simulation of magnetic Hamiltonians
\cite{glaetzle15,bijnen15}.  As we discuss more in detail below, LMG
can also appear as a coarse-grained model for electric or magnetic
quantum dipoles \cite{lahaye08}.

In the present work, we assume that the components of the LMG system
in interaction with a blackbody radiation are actual spins, like in
the ferromagnetic compounds, whereas trapped ions and ultracold
condensates, even if they behave as effective spins, can interact with
a blackbody radiation via other degrees of freedom.

The paper is organized as follows.  In Sec. \ref{LE}, we briefly
describe our LBA approach to thermalization.  The LBA scheme is then
specialized in Sec. \ref{BB}, where the environment is chosen to be a
black-body radiation. In Sec. \ref{LMGS}, we recall the definition of
the LMG model. In Sec. \ref{PR}, we investigate, under which
conditions, a description via a LMG model of spins interacting with a
black-body environment is correct, and when the fully coherent limit
is valid or not, via tuning of the particle density.  In
Sec. \ref{SR}, we derive a simple selection rule that takes place when
the fully coherent limit is realized.  In Sec. \ref{TLD}, we analyze
the isoptropic LMG model. Here, we specialize to the fully coherent
limit, where the total angular momenta remains conserved, and derive
analytically all the elements necessary to evaluate the thermalization
times.  For the latter, we first provide simple analytical evaluations
of both the decoherence and dissipation times, which are then
confirmed in Sec. \ref{numerical}, where we provide a complete
numerical analysis, allowing also for a clear picture of the finite
size effects, particularly strong near the critical point.  Finally,
several crucial conclusions are drawn.

\section{Thermalization via Lindblad Equation}
\label{LE}
Let us consider a system described by a Hamiltonian operator $\bm{H}$
acting on a Hilbert space $\mathscr{H}$ of dimension $M$.  We assume
that the eigenproblem, $\bm{H} \ket{m} = E_m \ket{m}$, has discrete
nondegenerate eigenvalues and that the eigenstates $\{\ket{m}\}$ form
an orthonormal system in $\mathscr{H}$.  We arrange the eigenvalues in
ascending order $E_1< E_2 < \dots < E_M$.

In the following we briefly resume our recently proposed LBA to the
thermalization of many-body systems with nondegenerate spectra, which
allows for an unambiguous definition of the thermalization times, also
for compounds of, possibly equal, noninteracting systems~\cite{OPprl}.

The Lindblad equation (LE) represents the most general class of
evolution equations of the reduced density matrix operator
$\bm{\rho}(t)$ of a system interacting with an environment under the
assumptions that this evolution is a semigroup, preserves hermicity,
positivity, and the trace of $\bm{\rho}(t)$ at all times.  The generic LE
equation can be written as
\begin{align}
  \label{GEN_LIN}%
  \frac{\D \bm{\rho}}{\D t} = -\frac{\I}{\hbar}
  \left[\bm{H}',\bm{\rho}\right] +\sum_{\alpha} \left(
    \bm{L}_{\alpha}\bm{\rho}\bm{L}^{\dag}_{\alpha} -\frac{1}{2}\left\{
      \bm{L}^{\dag}_{\alpha}\bm{L}_{\alpha},\bm{\rho} \right\}
  \right).
\end{align}
In this equation, the coherent part of the evolution is represented by
the Hermitian operator $\bm{H}'$ which, in general, differs from the
isolated system Hamiltonian $\bm{H}$.  The Lindblad, or quantum jump,
operators $\bm{L}_\alpha$ are, for the moment, completely arbitrary
operators. Even their number is arbitrary but can always be reduced to
$M^2-1$.  If $\bm{H}$ has a nondegenerate spectrum, one can represent
the most general set of these operators by dyadic products of
eigenstates of $\bm{H}$, namely, $\ell_{m,n} \ket{m}\bra{n}$. The
meaning of the coefficients $\ell_{m,n}$ is obtained by further
developing the theory.  When it is imposed that the stationary
condition of the system coincides with the Gibbs state,
$\bm{\rho}_G\propto \exp(-\beta \bm{H})$, for a given inverse
temperature $\beta$, the Lindblad equation, projected onto the
eigenstates of $\bm{H}$, benefits from a decoupling between the $M$
diagonal terms, $\rho_{n,n}$, and the $M(M-1)$ off-diagonals terms,
$\rho_{m,n}$, $m\neq n$, and, furthermore, the latter terms are each
other decoupled.

\textit{Diagonal terms (Pauli Equation).}  The diagonal terms obey the
following master-equation
\begin{align}
  \label{Pauli}
  \frac{\D p_{m}(t)}{\D t} = \sum_{n } \left[ p_{n }(t) W_{n \to
      m}-p_{m}(t )W_{m \to n} \right],
\end{align}
where $W_{m \to n}=|\ell_{n,m}|^2$ is the rate probability by which,
due to the interaction with the environment, a transition
$|m\rangle\to|n\rangle$ occurs.  In the weak coupling limit, these
rates can be calculated by using the time-dependent perturbation
theory.  The above Pauli equation can be written in vectorial form as
follows
\begin{align}
  \label{0PPt}
  \frac{\D \bm{p}(t)}{\D t}= -\bm{A}\ \bm{p}(t),
\end{align}
where $p_n=\rho_{n,n}$ and
\begin{align}
  \label{0Amn}
  A_{m,n} = \left\{
    \begin{array}{ll}
      -W_{n \to m}, &\qquad m\neq n, \\ 
      \sum_{k \neq m} W_{m \to k}, &\qquad m= n.
    \end{array}
  \right.
\end{align}

\textit{Off-Diagonal terms (decoherences).}  The $M(M-1)$ elements
$\rho_{m,n}$, $m\neq n$, behave as normal modes which relax to zero
according to
\begin{align}
  \label{0rhomn}
  |\rho_{m ,n }(t)|=|\rho_{m ,n }(0)|e^{-t/\tau_{m,n}},
\end{align}
where
\begin{align}
  \label{taumn}
  \tau_{m,n} = \left[ \frac{1}{2}\sum_k \left(W_{m \to k}+W_{n \to
        k}\right) \right]^{-1}.
\end{align}
The environment is supposed to remain in its own thermal equilibrium
at inverse temperature $\beta$.  Mathematically, this information is
encoded in the fact that the matrix $W_{m \to n}$ is similar to a
symmetric matrix $C_{m,n}$ having non negative elements via the square
root of the Boltzmann factors $\exp(-\beta E_m)$ and $\exp(-\beta
E_n)$, namely,
\begin{align}
  \label{WC}
  e^{-\frac{\beta}{2} E_{m }} W_{m \to n} e^{\frac{\beta}{2} E_{n }} =
  C_{m ,n }.
\end{align}
If the transition rates $W_{m \to n}$ satisfy Eq.~(\ref{WC}) for some
matrix $C_{m,n}$ with $C_{m,n}=C_{n,m}\geq 0$, then the stationary
state of the LE coincides with the Gibbs state $\bm{\rho}_G$, i.e.,
the stationary solution of the Pauli Eq.~(\ref{Pauli}) is $p_m=e^{-\beta E_m}/Z$,
where $Z=\sum_k e^{-\beta E_k}$, and $\rho_{m,n}=0$, $m\neq n$.

The characteristic time $\tau$ by which the system reaches the
stationary state is thus due to two different processes:
\begin{subequations}
  \label{0tau}
  \begin{align}
    \label{01tau}
    &\tau=\max\left\{\tau^{(P)},\tau^{(Q)}\right\},
    &&\quad \mathrm{thermalization~time},\\
    \label{0tauP}
    &\tau^{(P)}=\frac{1}{\mu_2(\bm{A})},
    &&\quad \mathrm{dissipation~time},\\
    \label{0tauQ}
    &\tau^{(Q)}=\max_{m\neq n}\tau_{m,n}, &&\quad
    \mathrm{decoherence~time}.
  \end{align}
\end{subequations}
The matrix $\bm{A}$ associated to the Pauli Eq.~(\ref{Pauli}) has a
unique zero eigenvalue and $M-1$ positive eigenvalues~\cite{OP}.  In
the above definition of $\tau^{(P)}$, $\mu_2(\bm{A})$ is the smallest
nonzero eigenvalue of $\bm{A}$.  The natural interpretation of
$\tau^{(P)}$ is that it represents a characteristic time by which the
system looses or gains energy, whereas $\tau^{(Q)}$ represents a
characteristic time by which the system looses quantum coherence.

The above LBA satisfies a series of minimal physical requirements, as
evident when applied to the case in which the environment is a
blackbody radiation, which will be briefly illustrated in the next
Section. We stress that the remarkable simplicity of our equations is
not due to some heuristic approach: they originate uniquely from the
Lindblad class when the Gibbs stationary state is imposed.  The LBA is
equivalent to the popular quantum optical master equation
(QOME)~\cite{Petruccione}, only when there is no degeneracy in the
energy levels as well as in the energy gaps of $\bm{H}$~\cite{OPprl}.
As we shall show later, when we consider the subspace where the total
angular momenta $\bm{J}^2$ is fixed, in the LMG model the energy
levels as well as the energy gaps are nondegenerate (see
Eq.~(\ref{HISO})).  Therefore, in the subspace where $\bm{J}^2$ is
fixed, all the results that we obtain could be equally derived from
the QOME.

\section{Blackbody radiation}
\label{BB}
In the case in which the environment is a blackbody radiation at
inverse temperature $\beta$, the time-dependent perturbation theory
combined with the Planck-law yields (this result can be reached by
treating the electromagnetic field classically, provided at the end
the contribution due to the spontaneous emission is added)
\begin{align}
  \label{0Bm_explicit}
  A_{m,m}=& \sum_{k:\,E_k<E_m} D_{k ,m }
  \frac{(E_{m}-E_{k})^3}{1-e^{-\beta\left(E_{m}-E_{k}\right)}}
  \nonumber \\ &+ \sum_{k:\,E_k>E_m} D_{k ,m }
  \frac{(E_{k}-E_{m})^3}{e^{\beta\left(E_{k}-E_{m}\right)}-1},
\end{align}
whereas the off-diagonal terms $m\neq n$ of $\bm{A}$ are
\begin{align}
  \label{0Amn_explicit}
  A_{m,n} = \left\{
    \begin{array}{ll}
      -D_{m,n}\frac{\left(E_{m}-E_{n}\right)^3}
      {e^{\beta\left(E_{m}-E_{n}\right)}-1}, &\qquad E_m>E_n,
      \\ 
      0, &\qquad E_m=E_n, 
      \\
      -D_{m,n}\frac{\left(E_{n}-E_{m}\right)^3}
      {1-e^{-\beta\left(E_{n}-E_{m}\right)}}, &\qquad E_m<E_n.
    \end{array}
  \right.
\end{align}
The coefficients $D_{m,n}$ are magnetic or electric dipole matrix
elements, whose value depend on the properties of the system embedded
in the blackbody radiation as follows.  In the following, we focus on
the case in which the system interacts with the electromagnetic (EM)
field through magnetic dipole operators
$\mu\bm{\sigma}_i=(\mu\sigma_i^x,\mu\sigma_i^y,\mu\sigma_i^z)$, where
the index $i$ labels the individual elements of the system located at
position $\bm{r}_i$.  All the dynamics is encoded in the internal
degrees of freedom, therefore all the particles are considered fixed
in space.  Based on the analysis of \cite{FGR} the thermalization
dynamics is characterized by three regimes:
\\
\textit{Fully coherent regime.}~ If the following condition holds
\begin{align}
  \label{coherent}
  \left|E_n-E_m\right| \ll hc/\ell, \qquad \ell=\max_{i\neq j}
  \left|\bm{r}_i-\bm{r}_j\right|,
\end{align}
then the following formula applies
\begin{align}
  \label{Dnm.coherent}
  D_{n,m} &= \gamma\sum_{h=x,y,z} \left|\bra{n} \sum_{i=1}^{N}
    \sigma_i^{h} \ket{m}\right|^2 ,
\end{align}
where the coupling constant $\gamma$, in Gaussian units, can be
expressed in terms of the magnetic dipole operator and fundamental
constants as:
\begin{align}
  \label{gamma}
  \gamma=\frac{4\mu^2}{3\hbar^4 c^3}.
\end{align}

For $N=1$, Eq.~(\ref{Dnm.coherent}) equals the standard textbook
formula based on the long wavelength approximation \cite{DAV}.
\\
\textit{Fully incoherent regime.}~ If the following condition holds
\begin{align}
  \label{incoherent}
  \left|E_n-E_m\right| \gg hc/a, \qquad a=\min_{i\neq j}
  \left|\bm{r}_i-\bm{r}_j\right|.
\end{align}
then the following formula applies
\begin{align}
  \label{Dnm.incoherent}
  D_{n,m} &= \gamma \sum_{i=1}^{N} \sum_{h=x,y,z} \left|\bra{n}
    \sigma_i^{h} \ket{m}\right|^2 .
\end{align}
Since $hc = 1.23\ \mbox{eV $\mu$m}$, we have that for atomic or
molecular systems in which $|E_n-E_m|$ is typically of a few eV and
$\ell$ is not larger than a few tens of \AA,
condition~(\ref{coherent}) is well satisfied.  Instead, for
microscopic systems in which $a$ is $1~\mu\mbox{m}$ and the
energy-level separations $|E_n-E_m|$ are much larger than the atomic
eV scale, condition (\ref{incoherent}) applies.

Concerning the incoherent limit, from Eqs.~(\ref{0Bm_explicit}) and
(\ref{0Amn_explicit}) we see that, even if for some pairs of states
$\ket{m},\ket{n}$, the condition~(\ref{incoherent}) is not satisfied,
the contribution corresponding to such pairs can be neglected if
$\beta\Delta E \ll 1$, where $\Delta E$ is the largest of the values
$|E_n-E_m|$ for which the condition (\ref{incoherent}) does not hold.
From Eq.~(\ref{incoherent}) we see that a sufficient condition for
this to occur is
\begin{align}
  \label{incobeta}
  \beta hc/a \ll 1.
\end{align}
\\
\textit{Intermediate regime.}~ When none of the above inequalities
(\ref{coherent}), (\ref{incoherent}) and (\ref{incobeta}) hold, there
is no simple formula to be applied, and one should include
contributions with mixed dipole matrix elements. These contributions
originate from the general formula for the transition probabilities of
a many-body system perturbed by the presence of the black-body
radiation \cite{FGR}:
\begin{align}
  \label{Pnm.general}
  P^\pm_{n,m} &= \frac{\mu^2}{2\pi\hbar c^3}\
  \frac{\omega_{n,m}^3}{e^{\hbar\omega_{n,m}/k_B T}-1} \nonumber \\
  &\quad\times \sum_{i=1}^{N} \sum_{j=1}^{N} \sum_{h=1}^{3}
  \sum_{l=1}^{3} Q_{n,m}^{i,j;h,l} \nonumber \\ &\quad\times \bra{E_n}
  \sigma_i^{h} \ket{E_m} \overline{\bra{E_n} \sigma_j^{l} \ket{E_m}} ,
\end{align}
where
\begin{align}
  \label{omeganm}
  \omega_{n,m}=|E_n-E_m|/\hbar,
\end{align}
and
\begin{align}
  \label{Qnm}
  Q_{n,m}^{i,j;h,l} = \int_{0}^{\pi} \!\!  \sin\theta \D{\theta}
  \int_{0}^{2\pi} \!\!\! \D{\phi}\
  e^{\I\bm{u}\cdot(\bm{r}_i-\bm{r}_j)\omega_{n,m}/c}
  \left(\delta_{h,l}-u_hu_l\right).
\end{align}
with $\bm{u}=(\sin\theta\cos\phi,\sin\theta\sin\phi,\cos\theta)$.
Equation~(\ref{Pnm.general}), interpolates between the fully coherent
and fully incoherent limits.  Later on, we shall make use of
Eq.~(\ref{Pnm.general}) to show that, in the LMG model, as soon as
condition~(\ref{coherent}) is not satisfied, $\bm{J}^2$ is not
conserved.

\section{The Lipkin-Meshkov-Glick model}
\label{LMGS}
Let us consider the Hilbert space $\mathcal{H}$ of $N$ spins $\bm{S} =
\bm{\sigma} \hbar/2$, where $\bm{\sigma}=(\sigma^x,\sigma^y,\sigma^z)$
are the standard Pauli matrices. The dimension of $\mathcal{H}$ is
$M=2^N$.  The LMG model is defined in $\mathcal{H}$ through the
Hamiltonian
\begin{align}
  \label{LMG}
  \bm{H}=-\frac{\mathcal{J} \hbar^2}{4N} \sum_{i\neq j}^{N}
  \left(\sigma^{x}_i\sigma^{x}_j+\gamma_y\sigma^{y}_i\sigma^{y}_j\right)-
  \frac{\Gamma \hbar}{2}\sum_{i=1}^{N} \sigma^{z}_i,
\end{align}
where $\mathcal{J}$ is the spin-spin coupling, $\Gamma$ the strength
of a transverse field, and $\gamma_y$ the so called anisotropy
parameter.  The model is known to provide an exactly solvable
mean-field like behavior in the limit $N\to\infty$~\cite{LMG}.  Let us
introduce the components $h=x,y,z$ of the total spin operator $\bm{J}$
\begin{align}
  \label{LMG1}
  J_h=\frac{\hbar}{2}\sum_{i=1}^{N} \sigma^{h}_i.
\end{align}
Up to the additive constant $\mathcal{J}\hbar^2(1+\gamma_y)/4$, we can
rewrite the Hamiltonian as
\begin{align}
  \label{LMG2}
  \bm{H}=-\frac{\mathcal{J}}{N}\left(J_x^2 +\gamma_y
    J_y^2\right)-\Gamma J_z.
\end{align}
It follows that $[\bm{H},\bm{J}^2]=0$ and
$[\bm{H},\prod_i\sigma_i^z]=0$.  These two relations imply, whenever
the system of $N$ spins is isolated, the conservation of the total
spin $\bm{J}^2$, and the conservation of the parity along the $z$
direction.  As a consequence, the eigenstates $|m\rangle$ of $\bm{H}$
can be chosen as (the label $m$ here should not be confused with the
eigenvalues of $J_z$, for which we shall use the symbol $m_z$)
\begin{eqnarray}
  \label{LMG3}
  | m \rangle=|j,p,\alpha\rangle \in \mathcal{H}_j\cap\mathcal{H}^{(p)},
\end{eqnarray}
where $j$ is the quantum number associated to $\bm{J}^2$, i.e.,
$\bm{J}^2|j,p,\alpha\rangle= \hbar^2 j(j+1)|j,p,\alpha\rangle$, and $p=\pm 1$
is the parity, i.e., $\prod_i\sigma_i^z \ket{j,p,\alpha}= p
\ket{j,p,\alpha}$.  The Greek symbol $\alpha$ stands for a suitable
set of quantum numbers that allow the state $|j,p,\alpha\rangle$ to
span the intersection between the $2j+1$ dimensional Hilbert space
$\mathcal{H}_j$, where $j$ is fixed, and the $2^N/2$ dimensional
Hilbert space $\mathcal{H}^{(p)}$, known as the ``half space'' of
$\mathcal{H}$ in which $p$ is fixed.  According to the rules for the
addition of angular momenta, for $N$ spins 1/2 we have
\begin{subequations}
  \label{jrange}
  \begin{align}
    & N~\mathrm{odd}\Rightarrow j\in\{1/2,3/2,\ldots,N/2\}, \\
    & N~\mathrm{even}\Rightarrow j\in\{0,1,\ldots N/2\}.
  \end{align}
\end{subequations}

In the symmetric case, $\gamma_y=1$, we also have $[\bm{H},J_z]=0$,
and the index pair $(p,\alpha)$ coincides with $(p,m_z)$, where $m_z$
is the eigenvalue of $J_z/\hbar$, taking the values $-j,-j+1,\ldots,j$
restricted to either $p=1$ or $p=-1$ (if two values $m_z$ and $m_z'$
have the same parity, then $|m_z'-m_z|$ can be either 0 or 2).

\section{Implementations of LMG with magnetic systems in a blackbody
  environment}
\label{PR}
In this Section, we want to analyze which regime, fully coherent,
fully incoherent, or intermediate, takes place in realistic models
characterized by an effective LMG description.  We restrict to two
magnetic systems with permanent magnetic moment.  A more general
analysis devoted to the study of atomic/molecular systems with
electric dipole moments will be done somewhere else.

In general, the conditions (\ref{coherent}) or (\ref{incoherent}),
must be checked for all those pairs $(m,n)$ of eigenstates
contributing with non zero dipole elements (\ref{Dnm.coherent}) and
(\ref{Dnm.incoherent}).  However, in the LMG model, as well as in
models characterized by a smooth energy landscape, near states
correspond to near energies and, moreover, since the operators
associated to the dipole matrix elements are sums of Pauli matrices,
the dipole matrix elements can connect only states that differ by
single spin flips.  Therefore, the pairs $(m,n)$ for which we have to
control the conditions (\ref{coherent}) or (\ref{incoherent}), with
respect to possible dependencies on $N$, always have $|E_n-E_m|\sim
\mathcal{J} \hbar^2$.

The first realistic model of interest is provided by the so called
high-spin molecules. These are large molecules having a large total
spin $j$ (which defines the eigenvalues of $\bf{J}^2$), well described
by the LMG Hamiltonian~(\ref{LMG2}).  According to Ref.~\cite{Ziolo},
in the high-spin molecule Mn$_{12}$, we have $j=10$ and
$hc/(\mathcal{J}\hbar^2) \simeq 2~\mathrm{cm}$.  Substituting the latter value
in Eq.~(\ref{coherent}), we see that the fully coherent condition
becomes $\ell\ll 2~\mathrm{cm}$, which is certainly satisfied (the
diameter of the molecule cannot overcome a few tens of
\r{A}ngstr\"om).

The other class of realistic models, concerns magnetic ions in a
crystalline environment, such as $\mathrm{Dy(C_2H_5SO_4)_39H_2O}$ and
$\mathrm{DyPO_4}$ among others~\cite{Abragam,Wolf} and ultracold atoms
with a permanent dipole moment~\cite{lahaye08}.  Here, the
dipole-dipole interaction decays with the cube of the distance between
two neighboring ions and is anisotropic.  As a consequence, unless the
temperature is sufficiently high, as prescribed by
Eq.~(\ref{incobeta}), there is no way to stay in the fully incoherent
regime. This becomes clear by the following argument.  Two neighboring
spins $S_i$ and $S_j$ interact via the dipole-dipole Hamiltonian:
\begin{align}
  \label{Hmm}
  H_{i,j}={\displaystyle -{\frac {\mu _{0}}{4\pi |{\mathbf
          {r}}|^{3}}}\left[3({\mathbf {m}}_{1}\cdot {\mathbf
        {r}})({\mathbf {m}}_{2}\cdot {\mathbf {r}}){\dfrac
        {1}{|{\mathbf {r}}|^{2}}}-{\mathbf {m}}_{1}\cdot {\mathbf
        {m}}_{2}\right]},
\end{align}
where $\mu_0$ is the vacuum permeability, $\mathbf {m}_{1}$ and
$\mathbf {m}_{2}$ the magnetic moments of the two spins, and $|\mathbf
{r}|=a$ their distance.  Eq.~(\ref{Hmm}) allows to estimate the
coupling constant $\mathcal{J}$ in a coarse-grained Ising-like
Hamiltonian for spin $1/2$ particles $H=-\sum_{(i,j)}\mathcal{J} S_i
S_j$.  In fact, if each magnetic moment has an electronic origin, we
have $|\mathbf {m}_{i}|\sim \mu_B$, where $\mu_B$ is the Bohr
magneton.  By comparison between $\mathcal{J} S_i S_j$ and $H_{i,j}$,
we can rewrite the Ising coupling in terms of fundamental constants as
\begin{align}
  \label{JEST}
  \mathcal{J}\hbar^2 \sim \alpha^3 a_0^2 \frac{\pi\hbar c}{a^3},
\end{align}
where $\alpha_0$ it the fine~structure~constant, and $a_0$ is the
Bohr~radius.  We can now apply Eq.~(\ref{JEST}) to condition
(\ref{coherent}) and find that the fully coherent condition amounts
to:
\begin{align}
  \label{APPFC}
  \alpha^3 a_0^2 \ll \frac{a^3}{\ell},
\end{align}
while applying Eq.~(\ref{JEST}) to condition (\ref{incoherent}) we see
that the fully incoherent condition amounts to:
\begin{align}
  \label{APPFI}
  \alpha^3 a_0^2 \gg a^2.
\end{align}
Since $a\geq a_0$, and $\alpha\simeq 1/137$, we see that
Eq.~(\ref{APPFI}) is never satisfied.  Eq.~(\ref{APPFC}) can be
instead satisfied at sufficiently low densities.  In fact, since
$\ell\sim a_0 N^{1/d}$, where $d$ is the dimension (real or effective)
of the system, we see that Eq.~(\ref{APPFC}) is satisfied if $a$ grows
with $N$ at least as $a\sim a_0 N^{1/(2d)}$. Whereas for finite $d$ such a
condition amounts, in the thermodynamic limit, to infinitely small
densities, for $d=\infty$, like occurs in a fully connected model,
Eq.~(\ref{APPFC}) is certainly satisfied for any finite density.
However, the fully connected interaction is a theoretical
extrapolation, as the actual $d$ will remain finite. In this sense, we
can consider the LMG model as a mean-field approximation of the finite
dimensional case.  As a consequence, we expect that some trade-off
will take place, with the fully coherent limit being satisfied only
for densities lower than some threshold.  The numerical value of this
threshold could be calculated on the base of the specific limiting
procedure $d\to\infty$ chosen for defining the LMG model, which is
beyond the aim of the present work.  In any case, a threshold exists,
and for densities higher than the threshold, neither Eq.~(\ref{APPFC})
and nor Eq.~(\ref{APPFI}) are satisfied, and the general formula
(\ref{Pnm.general}) should be applied instead.  In the next Section,
we will make use of the general formula (\ref{Pnm.general}) to show
that, along the thermalization, whenever the fully coherent regime is
not satisfied, as occurs at high densities, the total angular momenta
is not conserved, while, for the rest of the paper, we will perform a
comprehensive analysis of the thermalization by assuming the fully
coherent regime, as is expected to take place at low densities.

\section{Selection rules for the thermalization of the LMG model}
\label{SR}
To determine the matrix elements of $\bm{A}$ from
Eqs.~(\ref{0Bm_explicit}) and (\ref{0Amn_explicit}), one must evaluate
the dipole matrix elements $D_{m,n}$.  Let us indicate by
$|m\rangle=|j,p,\alpha\rangle$ and $|n\rangle=|j',p'\alpha'\rangle$,
two generic eigenstates of $\bm{H}$.  Since $[\bm{J}^2,J_h]=0$, for
any $h=x,~y,~z$, we clearly have
\begin{align}
  \label{JJ}
  J_h|m\rangle =J_h|j,p,\alpha\rangle \in \mathcal{H}_j,
\end{align}
so that, if we assume the fully coherent regime, from
Eq.~(\ref{Dnm.coherent}), for dipole matrix element we have
\begin{align}
  \label{FC}
  D_{m;n} &= D_{j,p,\alpha;j',p',\alpha'} = 0 \qquad \mbox{if $j\neq
    j'$}.
\end{align}
Furthermore, whereas $J_z$ conserves the parity, this is not true for
$J_x$ and $J_y$, so that, in general, we also have
\begin{align}
  \label{FC1}
  D_{m;n} &= D_{j,p,\alpha;j,p',\alpha'} \neq 0.
\end{align}

Let us consider now the fully incoherent regime.  Consider, for
example, the symmetric case $\gamma_y=1$, where
$|j,p,\alpha\rangle=|j,p,m_z\rangle$ and choose $N=2$.  The basis is
spanned by the singlet state $j=0$, $m_z=0$ and the triplet states
$j=1$, $m_z=-1,0,+1$.  From Eq.~(\ref{Dnm.incoherent}), we see that
the dipole matrix elements contain, for example, contributions
proportional to:
\begin{align}
  \label{ARGO}
  &|\langle j=1,p,m_z=1|\sigma^z_1|j=0,p',m_z=0\rangle| = 0, \\
  &|\langle j=1,p,m_z=1|\sigma^x_1|j=0,p',m_z=0\rangle| = \nonumber \\
  \label{ARGOb}
  &|\langle j=1,p,m_z=1|\sigma^y_1|j=0,p',m_z=0\rangle|\neq 0,
\end{align}
(and similarly for $\sigma_2^h$, $h=x,~y,~z$) which give
\begin{align}
  \label{FI}
  D_{m;n} &= D_{j,p,\alpha;j',p',\alpha'} \neq 0, \qquad \mbox{even if
    $j \neq j'$}.
\end{align}

Finally, let us consider the intermediate regime, and for simplicity
let us again consider a system with $N=2$. From the rhs of the general
formula (\ref{Pnm.general}), we see that, in particular, the
contributions corresponding to the case $i=j$ and $h=l$, are
proportional to the terms (\ref{ARGO})-(\ref{ARGOb}) and alike.

Eqs.~(\ref{FC}), (\ref{FC1}), and (\ref{FI}), show that, whereas the
thermalization process is always able to connect states with different
parity, in the fully coherent regime the thermalization process does
not connect states with different total spin, whereas it is able to do
so outside of this regime.

In the following, we will disregard the description of the states in
terms of $p$ and we shall use the notation $|j,\alpha\rangle$ since,
regardless of the regime, the parity of the state does not provide any
useful selection rule.

\section{Thermalization in the fully coherent regime for isotropic LMG
  models}
\label{TLD}
In the fully coherent limit, if the system is initially prepared in a
mixture, $\bm{\rho}_j(t=0)$, of eigenstates of $\bm{J}^2$, all with
eigenvalues $j$, it will remain in the subspace $\mathcal{H}_j$ for
all times. In other words, the system will undergo a partial
thermalization, reaching asymptotically the following thermal state
\begin{align}
  \label{TS}
  \lim_{t\to\infty}\bm{\rho}_j(t)=\frac{\exp(-\beta
    \bm{H})\bm{P}_j}{Z_j},
\end{align}
where $\bm{P}_j$ is the projector onto $\mathcal{H}_j$, and
$Z_j=\mathrm{tr}(\exp(-\beta\bm{H})\bm{P}_j)$.

We now briefly review the properties of the isotropic LMG model and
discuss in detail its thermalization properties.  
If $\gamma_y=1$, the
Hamiltonian (\ref{LMG2}) simplifies as
\begin{align}
  \label{HISOa}
  \bm{H}=-\frac{\mathcal{J}}{N}J^2 +\frac{\mathcal{J}}{N}J_z^2 -\Gamma
  J_z,
\end{align}
where, as long as we are confined in the subspace $\mathcal{H}_j$, the
first term is a constant.  Note that, whereas in the full Hilbert
space $\mathcal{H}$ the Hamiltonian Eq.~(\ref{HISOa}) leads to a
ferromagnetic phase, in $\mathcal{H}_j$, if $\mathcal{J}>0$, as
usually assumed in the LMG models, Eq.~(\ref{HISOa}) represents the
classical Hamiltonian of a fully connected Ising model with an
anti-ferromagnetic coupling, a highly frustrated system with no
ordinary finite temperature phase-transition.  Therefore, a phase
transition can occur in the LMG model only at zero temperature, and
the order parameter must be properly defined~\cite{LMG_Botet}.  In
order to have some magnetization in $\mathcal{H}_j$ with a finite
temperature phase-transition, one must allow $\gamma_y$ to be
different from $1$.  An explicit classical analysis of the finite
temperature phase transition can be found in~\cite{LMG_Das}.  We
stress that, even if, for $\gamma_y=1$, the Hamiltonian~(\ref{LMG2})
is somehow classical, its thermalization is governed by genuine
quantum processes.  More precisely, the interaction with the
surrounding EM field is not trivial since all the three components of
the total spin participate.

Below we provide an exact analysis of the thermalization of the LMG
model for $\gamma_y=1$.  We first analyze the static and equilibrium
properties, and then calculate the dipole matrix elements which, in
turn, allow us to evaluate the thermalization times by using the
equations discussed in Secs. II and III.

\subsection{Energy levels, gap, and critical point}
In the following we will work in units where $\hbar=1$.
If $\gamma_y=1$, the eigenstates of the Hamiltonian $\bm{H}$ are
simply given by $|m\rangle=|j,m_z\rangle$, with eigenvalues
\begin{align}
  \label{HISO}
  E(j,m_z)=-\frac{\mathcal{J}
    j(j+1)}{N}+m_z\left(\frac{\mathcal{J}m_z}{N}-\Gamma\right),
\end{align}
with $m_z\in\{-j,-j+1\ldots,j\}$. We assume $N\geq 2$. Furthermore, we
consider $j>0$, otherwise there exists only one state
$\ket{j=0,m_z=0}$.  As a consequence, we have $j\geq 1$ integer if $N$
is even or semi-integer if $N$ is odd.

From Eq.~(\ref{HISO}) we have (hereafter, since $j$ is fixed, we use
the shorter notation $E_{m_z}=E(j,m_z)$)
\begin{subequations}
  \label{DEISO}
  \begin{align}
    & E_{m_z}-E_{m_z-1}=\frac{(2m_z-1) \mathcal{J}-\Gamma N}{N},
    \qquad m_z\geq -j+1,\\
    & E_{m_z}-E_{m_z+1}=\frac{\Gamma N-(2m_z+1)\mathcal{J}}{N}, \qquad
    m_z\leq j-1.
  \end{align}
\end{subequations}
In the following, we indicate by $m^{(1)}_z$ the ground state (GS),
and by $m^{(2)}_z$ the first excited state (FES).  Let us suppose for
the moment being that $\Gamma N/(2\mathcal{J})$ is not an half integer
for $j$ even (is not an integer for $j$ odd) so that, even for $N$
finite, the gaps do not close.  For the GS, we have
\begin{widetext}
  \begin{align}
    \label{GSISO}
    E_{m^{(1)}_z}=\min_{m_z} E_{m_z}, \qquad
    m^{(1)}_z=\mathrm{sgn}(\Gamma)
    \min\left\{\left[\frac{|\Gamma|N}{2\mathcal{J}}\right]_j,j\right\},
  \end{align}
  where we have defined
  \begin{align}
    \label{NINT} [x]_j= \left\{
      \begin{array}{ll}
        \mbox{integer closest to $x$}, &\qquad\mbox{$j$ even}, \\
        \mbox{semi-integer closest to $x$}, &\qquad\mbox{$j$ odd}.
      \end{array}
    \right.
  \end{align}
  It is convenient to introduce
  \begin{align}
    \label{deltaISO}
    \delta=\left[\frac{\Gamma N}{2\mathcal{J}}\right]_j-\frac{\Gamma
      N}{2\mathcal{J}}.
  \end{align}
  By using Eqs.~(\ref{HISO}) and (\ref{GSISO}), and the definition of
  $\delta$, for the GS and FES levels we obtain
  \begin{align}
    \label{GSISOEXPL}
    E_{m^{(1)}_z}=\left\{
      \begin{aligned}
        &-\frac{\mathcal{J} j(j+1)}{N}-\frac{\Gamma^2 N}
        {4\mathcal{J}}+\frac{\mathcal{J}\delta^2}{N}, &&\qquad
        \Gamma/\mathcal{J}\in \left[-\frac{2(j-\delta)}{N},
          \frac{2(j-\delta)}{N}\right], \\
        &-\frac{\mathcal{J}
          j(j+1)}{N}+\frac{\mathcal{J}j^2}{N}-j|\Gamma|, &&\qquad
        \Gamma/\mathcal{J}\notin
        \left[-\frac{2(j-\delta)}{N},\frac{2(j-\delta)}{N}\right],
      \end{aligned}
    \right.
  \end{align}
  \begin{align}
    \label{FELISO}
    E_{m^{(2)}_z}=\min_{m_z\neq m^{(1)}_z} E_{m_z}=\left\{
      \begin{aligned}
        &E_{m^{(1)}_z-\mathrm{sgn}(\delta)},
        &&\qquad |m^{(1)}_z-\mathrm{sgn}(\delta)|\leq j, \\
        &E_{m^{(1)}_z+\mathrm{sgn}(\delta)}, &&\qquad
        |m^{(1)}_z-\mathrm{sgn}(\delta)|>
        j~\mathrm{and}~|m^{(1)}_z+\mathrm{sgn}(\delta)|\leq j, \\
        &E_{\mathrm{sgn}(\Gamma)(j-1)}, &&\qquad
        m^{(1)}_z=\mathrm{sgn}(\Gamma)j.
      \end{aligned}
    \right.
  \end{align}
  From Eqs.~(\ref{DEISO})-(\ref{FELISO}) we evaluate the first gap
  $\Delta$
  \begin{align}
    \label{GAPISO}
    \Delta=E_{m^{(2)}_z}-E_{m^{(1)}_z}=\left\{
      \begin{aligned}
        &|\Gamma|-\mathcal{J}\frac{2j-1}{N}, &&\qquad
        \frac{\Gamma}{\mathcal{J}}\notin
        \left[-\frac{2(j-\delta)}{N},\frac{2(j-\delta)}{N}\right], \\
        &\mathcal{J}\frac{1+2|\delta|}{N}, &&\qquad
        \frac{\Gamma}{\mathcal{J}}\in
        \left[-\frac{2(j-\delta)}{N},-\frac{2(j-r(\delta)-\delta)}{N}\right]
        \cup
        \left[\frac{2(j-r(\delta)-\delta)}{N},\frac{2(j-\delta)}{N}\right],\\
        &\mathcal{J}\frac{1-2|\delta|}{N}, &&\qquad
        \frac{\Gamma}{\mathcal{J}}\in
        \left[-\frac{2(j-r(\delta)-\delta)}{N},
          \frac{2(j-r(\delta)-\delta)}{N}\right],
      \end{aligned}
    \right.
  \end{align}
\end{widetext}
where $r(\delta) =1$ if $\delta\cdot\Gamma <0$ and $r(\delta)=0$
otherwise.  If $r(\delta)=0$, the intermediate intervals in the second
line of Eq.~(\ref{GAPISO}) are empty sets.  Equation~(\ref{GAPISO})
shows that, for $N$ finite, we can define two ``exact critical
points'', ${\Gamma}_c^+$ and ${\Gamma}_c^-$, as solutions,
respectively, of the equations:
\begin{align}
  \label{Gammacex}
  \frac{\Gamma_c^{\pm}}{\mathcal{J}}=\pm 2\frac{(j-\delta)}{N}.
\end{align}
By using the definition of $\delta$, it is easy to check that, for any
$N$, Eqs.~(\ref{Gammacex}) are solved for $\Gamma$ such that
$\delta=0$, \textit{i.e.}
\begin{eqnarray}
  \label{Gammac}
  \frac{\Gamma_c^{\pm}}{\mathcal{J}}=
  \pm \frac{\Gamma_c}{\mathcal{J}}=\pm \frac{2j}{N}.
\end{eqnarray}
Notice that, for $j$ even (odd), the function $[x]_j$ has two values
for $x$ semi-integer (integer).  For $j$ even this reflects on the
fact that, whenever $\Gamma N/(2\mathcal{J})=k/2$, for some odd (even,
if $j$ is odd) integer $k$ such that $|k/2\pm 1/2|<j$, the GS level
can be two-fold degenerate, with the states $m^{(1a)}_z=k/2-1/2$ and
$m^{(1b)}_z=k/2+1/2$.  The general expression of the GS, as well as of
the FES, for the case in which $\Gamma N/(2\mathcal{J})$ is
semi-integer for $j$ even (or integer for $j$ odd) is cumbersome.  It
is however clear that such a condition on the external field $\Gamma$,
is of no physical interest, since one can approach an integer or a
semi-integer by an infinite sequence of real numbers that are neither
integer nor half-integer.

Equation~(\ref{GAPISO}) shows that there is an inner region in
$\Gamma$ where the gap closes to zero as $\Delta=
(1-2|\delta|)\mathcal{J}/N$, a paramagnetic external region where
$\Delta$ remains finite, and a transient region, whose size tends to
zero as $1/N$ and $\Delta= (1+2|\delta|)\mathcal{J}/N$.

Finally, we point out that analogous formulas hold for the successive
gaps. For example, for the difference between the third and the second
energy level, $\Delta'$, there is an interval in $\Gamma$ where
$\Delta'$ goes to zero as $1/N$, and, for $N$ large, such interval and
gap differ for negligible terms from, respectively, the interval and
gap between GS and FES.

\subsection{Partition function}
For later use, we also calculate the partition function $Z_j$ for $j$
large of the type $j=\alpha N$, with $\alpha$ constant.  From
Eq.~(\ref{HISO}) we have
\begin{align}
  \label{ZjISO}
  Z_j &= e^{\frac{\beta\mathcal{J} j(j+1)}{N}}
  \sum_{m_z\in[-j,-(j-1),\ldots,j]} e^{\beta N m_z
    \left(\frac{\mathcal{J}m_z}{N}-\Gamma\right)} \nonumber \\ &=
  e^{\frac{\beta\mathcal{J}
      j(j+1)}{N}}\sum_{x\in[-1,-(j-1)/j,\ldots,1]} e^{\beta \alpha x N
    \left(\mathcal{J}\alpha x-\Gamma\right)}.
\end{align}
For large $N$, the above sum can be approximated by an integral over
the range $(-1,1)$, and we get
\begin{align}
  \label{ZjISO1}
  Z_j=\sqrt{\frac{2\pi N}{\beta J \alpha^2}}e^{\frac{\beta\mathcal{J}
      j(j+1)}{N}}e^{\beta\Gamma \frac{\Gamma
      N}{2\mathcal{J}}}\left[1+O\left(\frac{1}{N}\right)\right].
\end{align}
Notice the absence of the constant $\alpha$ in the second exponential.

\subsection{Dipole matrix elements}
In order to evaluate the dipole matrix elements, we shall make use of
the ladder operators $J_{\pm}=J_x\pm\mathrm{i}J_y$. Let us consider
two generic eigenstates $|m\rangle=|j,m_z\rangle$ and
$|n\rangle=|j,n_z\rangle$, with $m_z,n_z\in\{-j,-j+1\ldots,j\}$. From
Eq.~(\ref{Dnm.coherent}), by using $D_{m,n}=\gamma\sum_{h}|\langle j,m_z|2
J^h|j,n_z\rangle|^2$, we have
\begin{widetext}
  \begin{align}
    \label{DISO}
    D_{m_z,n_z} = 2\gamma \left[(j-n_z)(j+n_z+1)\delta_{m_z,n_z+1}+
      (j+n_z)(j-n_z+1)\delta_{m_z,n_z-1}\right].
  \end{align}
  By plugging Eq.~(\ref{DISO}) into Eqs.~(\ref{0Bm_explicit}) and
  (\ref{0Amn_explicit}), with $A_{m,n} =A_{m_z,n_z}$, we get
  \begin{align}
    \label{BISO}
    A_{m_z,m_z} = 2\gamma(j-m_z+1)(j+m_z)f(E_{m_z-1}-E_{m_z})+
    2\gamma(j+m_z+1)(j-m_z)f(E_{m_z+1}-E_{m_z}),
  \end{align}
  and
  \begin{subequations}
    \label{AISO}
    \begin{align}
      & A_{m_z,m_z-1} = -2\gamma(j-m_z+1)(j+m_z)f(E_{m_z}-E_{m_z-1}), \\
      & A_{m_z,m_z+1}=-2\gamma(j+m_z+1)(j-m_z)f(E_{m_z}-E_{m_z+1}),  \\
      & A_{m_z,n_z}=0, \qquad n_z\neq m_z,~m_z-1,~m_z+1,
    \end{align}
  \end{subequations}
  where we have introduced the function $f(E_m)$:
  \begin{align}
    \label{fISO}
    f(E_{m_z}-E_{n_z})=
    \frac{(E_{m_z}-E_{n_z})^3}{e^{\beta\left(E_{m_z}-E_{n_z}\right)}-1}
    \theta(E_{m_z}-E_{n_z})+
    \frac{(E_{n_z}-E_{m_z})^3}{1-e^{-\beta\left(E_{n_z}-E_{m_z}\right)}}
    \theta(E_{n_z}-E_{m_z}),
  \end{align}
  $\theta(x)$ being the Heaviside step function.  Plugging
  Eqs.~(\ref{BISO}) into Eqs.~(\ref{0Amn}) and (\ref{taumn}), we
  calculate the decoherence times as
  \begin{align}
    \label{taumnISO}
    \tau_{m_z,n_z} =&\
    \left[2\gamma(j-m_z+1)(j+m_z)f(E_{m_z-1}-E_{m_z})+
      2\gamma(j+m_z+1)(j-m_z)f(E_{m_z+1}-E_{m_z})\right.  \nonumber \\
    &+ \left. 2\gamma(j-n_z+1)(j+n_z)f(E_{n_z-1}-E_{n_z})+
      2\gamma(j+n_z+1)(j-n_z)f(E_{n_z+1}-E_{n_z})\right]^{-1}, \qquad
    m_z\neq n_z.
  \end{align}
  In this framework, $j$ is fixed, but it can be chosen to be any
  value in agreement with Eqs.~(\ref{jrange}).  Notice that, since in
  Eq.~(\ref{taumnISO}) $m_z\neq n_z$, the values $j=0$ (for $N$ even)
  and $j=1/2$ (for $N$ odd), are not allowed (obviously, for such
  fixed values of $j$ we have no dynamics at all).  The decoherence
  times~(\ref{taumnISO}) can be easily evaluated numerically for any
  choice of the allowed $j$, $m_z$, and $n_z$.  Depending on the
  particular value of $\Gamma$, which determines the energy gap
  $\Delta$ via Eq.~(\ref{GAPISO}), we can have different
  thermalization regimes.  Below we provide analytical evaluations
  corroborated by exact numerical results.

  \subsection{Decoherence for $\Gamma/\mathcal{J}\notin
    \left[-\frac{2(j-\delta)}{N},\frac{2(j-\delta)}{N}\right]$}
  In this case, $\Delta$ is finite and, if $\beta\Gamma=O(1)$, from
  Eqs.~(\ref{DEISO}) we have
  \begin{align}
    \label{fISO1}
    f(E_{m_z\pm 1}-E_{m_z}) \sim O\left( \left|
        \Gamma-\mathcal{J}\frac{2m_z\pm 1}{N} \right|^3\right).
  \end{align}
  By using Eqs.~(\ref{fISO1}) in Eqs.~(\ref{taumnISO}), we get the two
  following possible scaling laws with respect to $j$
  \begin{subequations}
    \label{taumnISO1}
    \begin{align}
      &\tau_{m_z,n_z} = O\left(\frac{1}{\gamma \left|\Gamma\right|^3
          j^2}\right),
      \qquad |m_z|,~\mathrm{or}~|n_z| \ll j, \\
      &\tau_{m_z,n_z} = O\left(\frac{1}{\gamma \left|\Gamma-
            \mathcal{J}\frac{2j}{N}\right|^3 j}\right),
      \qquad m_z\sim n_z \sim j, \\
      &\tau_{m_z,n_z} = O\left(\frac{1}{\gamma
          \left||\Gamma|+\mathcal{J}\frac{2j}{N}\right|^3 j}\right),
      \qquad m_z\sim j,~ n_z \sim -j,
      \quad m_z\sim -j,~n_z\sim j, \\
      &\tau_{m_z,n_z} = O\left(\frac{1}{\gamma \left|\Gamma+
            \mathcal{J}\frac{2j}{N}\right|^3 j}\right), \qquad m_z\sim
      n_z \sim -j.
    \end{align}
  \end{subequations}
  Equations~(\ref{taumnISO1}) show that, for a given $j$, the states
  which remain coherent for a longer time are those with $m_z \sim n_z
  \sim \mathrm{sgn}(\Gamma) j$.  Quite importantly,
  Eqs.~(\ref{taumnISO1}) implies that, if $j$ is fixed and independent
  of $N$, the decoherence times do not scale with $N$ at all.
  Consider in particular the states with $j=0$. For $N$ even, these
  states are the sum of all the $N!$ permutations of spin-flips with
  alternate signs, i.e., the $N$-particle analogous of singlet
  2-particle state, an intrinsically entangled state.
  Equations~(\ref{taumnISO1}) tell us, if one is able to initially
  prepare the system with a small value of $j$, $N$-entangled states
  will show a strong resilience to decoherence. From the point of view
  of thermalization, this reflects on the overall thermalization time
  $\tau^{(Q)}$, which, from Eqs.~(\ref{taumnISO1}) becomes
  \begin{align}
    \label{tauQISO}
    \tau^{(Q)} = \max_{m_z\neq
      n_z}\tau_{m_z,n_z}=O\left(\frac{1}{\gamma
        \left||\Gamma|-\mathcal{J}\frac{2j}{N}\right|^3 j}\right).
  \end{align}
  In the limit of zero temperature $\beta\to\infty$, we can exploit
  \begin{align}
    \label{fbeta}
    \lim_{\beta\to \infty} f(E_{m_z}-E_{n_z})= \left\{
      \begin{aligned}
        &0, &&\qquad E_{m_z}> E_{n_z},
        \\
        &\left(E_{n_z}-E_{m_z}\right)^3, &&\qquad E_{m_z}<E_{n_z}.
      \end{aligned}
    \right.
  \end{align}
  By applying Eqs.~(\ref{fbeta}) and (\ref{taumnISO}), we achieve,
  roughly, the same overall behavior as Eq.~(\ref{tauQISO}).

  \subsection{Decoherence for $\Gamma/\mathcal{J}\in
    \left[-\frac{2(j-\delta)}{N},\frac{2(j-\delta)}{N}\right]$}
  In this case, $\Delta\sim \Delta'\sim \Delta''\ldots \sim 1/N$. If
  $\beta\mathcal{J}=O(1)$ and $\beta|\Gamma|=O(1)$,
  Eqs.~(\ref{FELISO}) and (\ref{GAPISO}) and their generalization for
  the successive gaps (whose details are not important here) show that
  \begin{align}
    \label{fISO2}
    f(E_{m_z\pm 1}-E_{m_z}) \sim
    O\left(\frac{|\Gamma|\mathcal{J}^2}{N^2}\right).
  \end{align}
  The interval in $\Gamma$ where Eq.~(\ref{fISO2}) can be applied to
  the arbitrary state $m_z$ is not trivial.  However, observing that
  Eq.~(\ref{fISO2}) can be applied to the GS and to the FES is enough
  to claim that, for $\Gamma/\mathcal{J}\in
  \left[-\frac{2(j-\delta)}{N},\frac{2(j-\delta)}{N}\right]$,
  \begin{align}
    \label{tauQISO2}
    \tau^{(Q)} = O\left(\frac{N^2 }{\gamma |\Gamma| \mathcal{J}^2
        \left(j^2+j-\left(\frac{\Gamma
              N}{2\mathcal{J}}\right)^2+\frac{|\Gamma|
            N}{2\mathcal{J}}\right)}\right),
  \end{align}
  where we have used Eq.~(\ref{GSISO}) for the explicit form of the
  GS.  From Eq.~(\ref{tauQISO2}) it follows that, if $j=O(N)$, then
  \begin{align}
    \label{tauQISO3}
    \tau^{(Q)}|_{\frac{|\Gamma|}{\mathcal{J}}= \frac{2j}{N} } =
    O\left(\frac{N }{\gamma |\Gamma|^2 \mathcal{J} }\right),
  \end{align}
  whereas
  \begin{align}
    \label{tauQISO4}
    \tau^{(Q)}|_{\frac{|\Gamma|}{\mathcal{J}}\ll \frac{2j}{N} } =
    O\left(\frac{1 }{\gamma |\Gamma| \mathcal{J}^2 }\right).
  \end{align}
  Equations~(\ref{tauQISO3}) and (\ref{tauQISO4}) show that, despite
  the gap closes to zero in all the interval
  $\left[-\frac{2j}{N},\frac{2j}{N}\right]$, the slow down dynamics
  takes place only in correspondence of the critical points
  $\Gamma_c^{\pm}/\mathcal{J}=\pm 2j/N$, and the decoherence time
  scales only linearly in $N$. On the other hand, we find remarkable
  to notice that, at the critical point, the decoherence time turns
  out to be a growing function of $N$.  This observation confirms and
  strengthen the general idea that phase transitions could be
  exploited to generate resilience to decoherence and large entangled
  states~\cite{Paganelli,Paganelli1}.

  Notice that Eq.~(\ref{fISO2}) is valid also for $\beta$ large,
  provided $N$ is sufficiently large too.  However, in general, the
  limits $\beta\to\infty$ and $N\to\infty$ cannot be switched.  If we
  are interested in
  $\lim_{N\to\infty}\lim_{\beta\to\infty}\tau_{m_z,n_z}$ we can simply
  use Eq.~(\ref{fbeta}) applied to Eq.~(\ref{taumnISO}). The special
  case at $\Gamma=\Gamma_c$ will be analyzed later.  If instead we are
  interested in $\lim_{\beta\to\infty}\lim_{N\to\infty}\tau_{m_z,n_z}$
  we can use Eqs.~(\ref{fISO2})-(\ref{tauQISO4}) by substituting
  everywhere one factor $|\Gamma|$ with $1/\beta$. This shows that, in
  the thermodynamic limit, the thermalization time diverges at least
  linearly in $\beta$.

  \subsection{Dissipation}
  In order to evaluate the dissipation time $\tau^{(P)}$, we must find
  the eigenvalue $\mu_2(\bm{A})$ of the $2j\times 2j$ matrix $\bm{A}$
  given in Eqs.~(\ref{BISO})-(\ref{AISO}). In general, this can be
  done only numerically. In the present case, this task is largely
  simplified because $\bm{A}$ is a tridiagonal matrix.

  From an analytical point of view, we can apply the general rule
  that, for $\beta$ finite, $\lim_{N\to\infty}\tau^{(P)}\geq
  \lim_{N\to\infty}\tau^{(Q)}$, with $\tau^{(Q)}$ given by
  Eqs.~(\ref{tauQISO}), (\ref{tauQISO3}), and (\ref{tauQISO4}).
  Eq. (\ref{tauQISO3}), in particular, implies that the thermalization
  time $\tau=\max\{\tau^{(P)},\tau^{(Q)}\}$, at the critical point and
  $\beta$ fixed diverges linearly in $N$.  Actually, the rule
  $\lim_{N\to\infty}\tau^{(P)}\geq \lim_{N\to\infty}\tau^{(Q)}$
  applies, if \cite{OP}
  \begin{align}
    \label{check}
    \lim_{N\to\infty}\frac{e^{-\beta E(j,m^{(1)}_z)}}{Z_j}=0.
  \end{align}
  Comparing Eq.~(\ref{GSISOEXPL}) with Eq.~(\ref{ZjISO1}), we see that
  the condition (\ref{check}) is verified for any value of $\Gamma$
  (with a decreasing factor that decays exponentially in $N$).  Notice
  that the inequality $\lim_{N\to\infty}\tau^{(P)}\geq
  \lim_{N\to\infty}\tau^{(Q)}$ holds for any $\beta$, so that we have
  also $\lim_{\beta\to\infty}\lim_{N\to\infty}\tau^{(P)}\geq
  \lim_{\beta\to\infty}\lim_{N\to\infty}\tau^{(Q)}$. However,
  $\lim_{N\to\infty}\lim_{\beta\to\infty}\tau^{(Q)}=2
  \lim_{N\to\infty}\lim_{\beta\to\infty}\tau^{(P)}$, since, in
  general, $\lim_{\beta\to\infty}\tau^{(Q)}=2
  \lim_{\beta\to\infty}\tau^{(P)}$~\cite{OP}.

  \subsection{Dissipation and decoherence at the critical point at
    zero temperature}
  The critical point at vanishing temperatures is intriguing.  Indeed,
  if we choose $N$ even and $j=N/2$, this setup coincides with the one
  used to investigate the quantum adiabatic
  algorithm~\cite{LMG_Santoro}.  From Eq.~(\ref{GAPISO}), for $N$
  large enough we have $\Gamma_c=\pm j$ and the GS is
  $m^{(1)}_z=\mathrm{sgn}(\Gamma)j$.  By using Eq.~(\ref{fbeta}), from
  Eqs.~(\ref{AISO}) we see that, for any finite $N$, in the limit
  $\beta\to\infty$ the matrix $\bm{A}$ becomes triangular and, as a
  consequence, from Eq.~(\ref{BISO}) for its lowest non zero
  eigenvalue $\mu_2$ we obtain
  \begin{align}
    \label{mu2crit}
    \lim_{\beta\to\infty}\mu_2(\bm{A})=2\gamma(2j-1)\Delta^3,
  \end{align}
  where $\Delta$ is given by Eq.~(\ref{GAPISO}) evaluated at
  $|\Gamma|\leq |\Gamma_c|=j$.  For $N$ large enough, we thus have:
  \begin{align}
    \label{tauPcrit}
    \lim_{\beta\to\infty}\tau^{(P)}=\frac{N^2}{2\gamma\mathcal{J}^3}.
  \end{align}
  Moreover, for the property $\lim_{\beta\to\infty}\tau^{(Q)}=2
  \lim_{\beta\to\infty}\tau^{(P)}$, we have also:
  \begin{align}
    \label{tauQcrit}
    \lim_{\beta\to\infty}\tau^{(Q)}=\frac{N^2}{\gamma\mathcal{J}^3},
  \end{align}
  and therefore:
  \begin{align}
    \label{taucrit}
    \lim_{\beta\to\infty}\tau=\frac{N^2}{\gamma\mathcal{J}^3}.
  \end{align}
  The present thermalization time $\tau$, which grows as $N^2$, is to
  be compared with the characteristic time to perform the quantum
  adiabatic algorithm~\cite{QAD}, which grows as $\tau_{ad} \sim
  N/\Delta^2=O(N^3)$.  This difference must be attributed to the
  spontaneous emission process, the only mechanism at $T=0$ by which
  the system, when in contact with the blackbody radiation, delivers
  its energy to the environment. Apparently, this mechanism provides a
  convergence toward the GS more efficient than that obtained in a
  slow transformation of the Hamiltonian parameters without
  dissipative effects.

  \begin{figure}
    \begin{center}
      {\includegraphics[width=0.45\textwidth,clip]{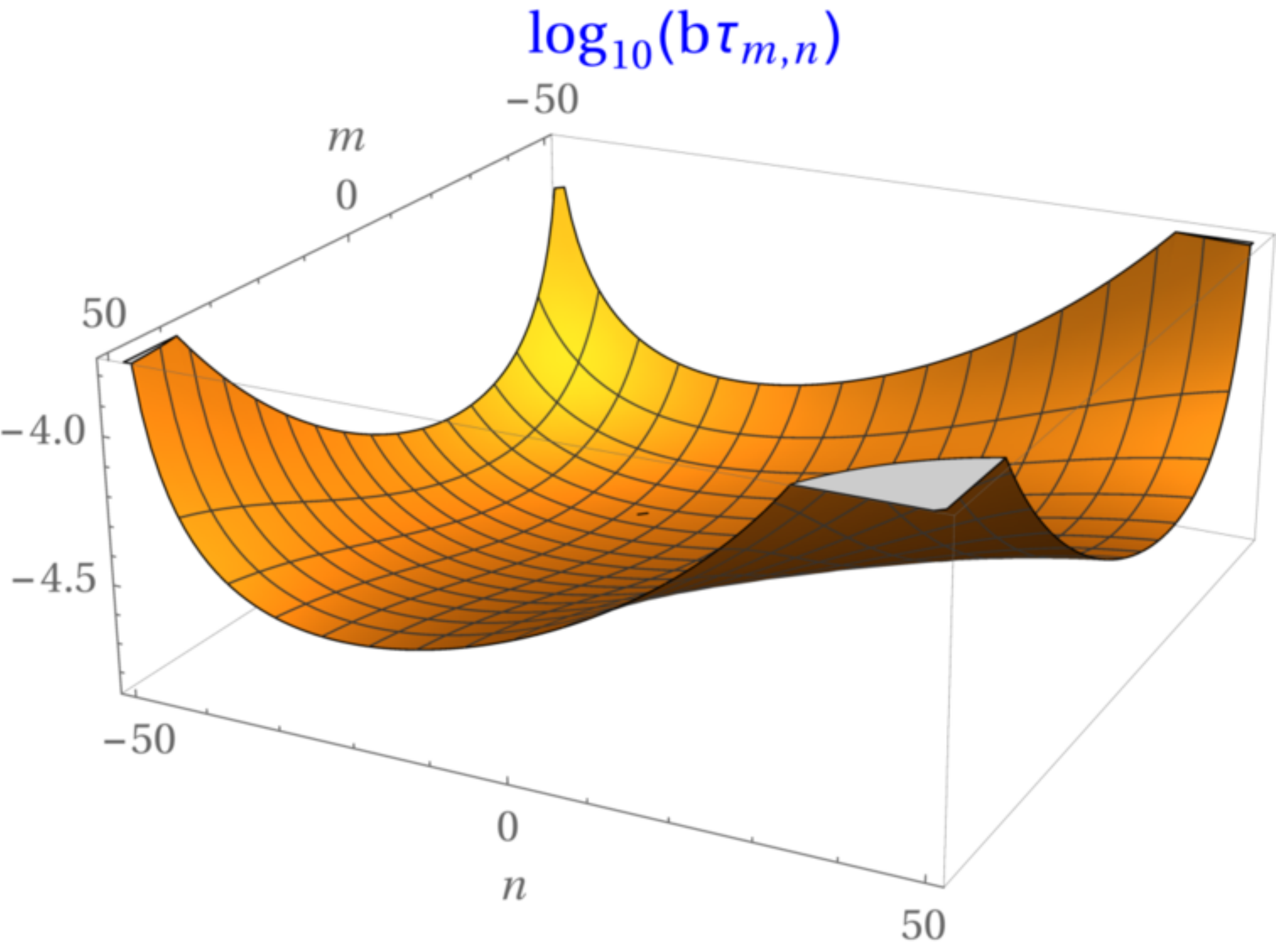}}
      {\includegraphics[width=0.45\textwidth,clip]{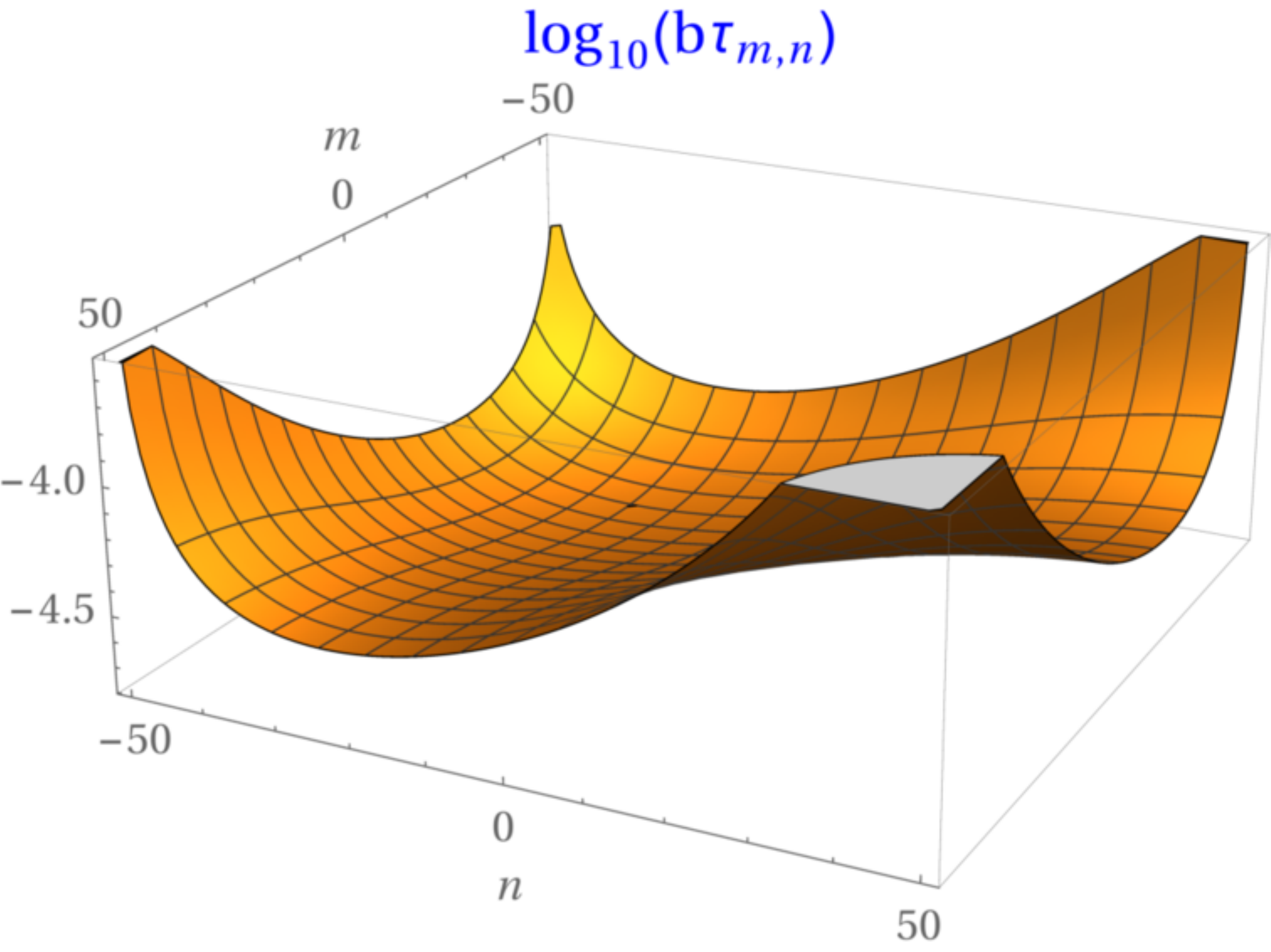}}
      {\includegraphics[width=0.45\textwidth,clip]{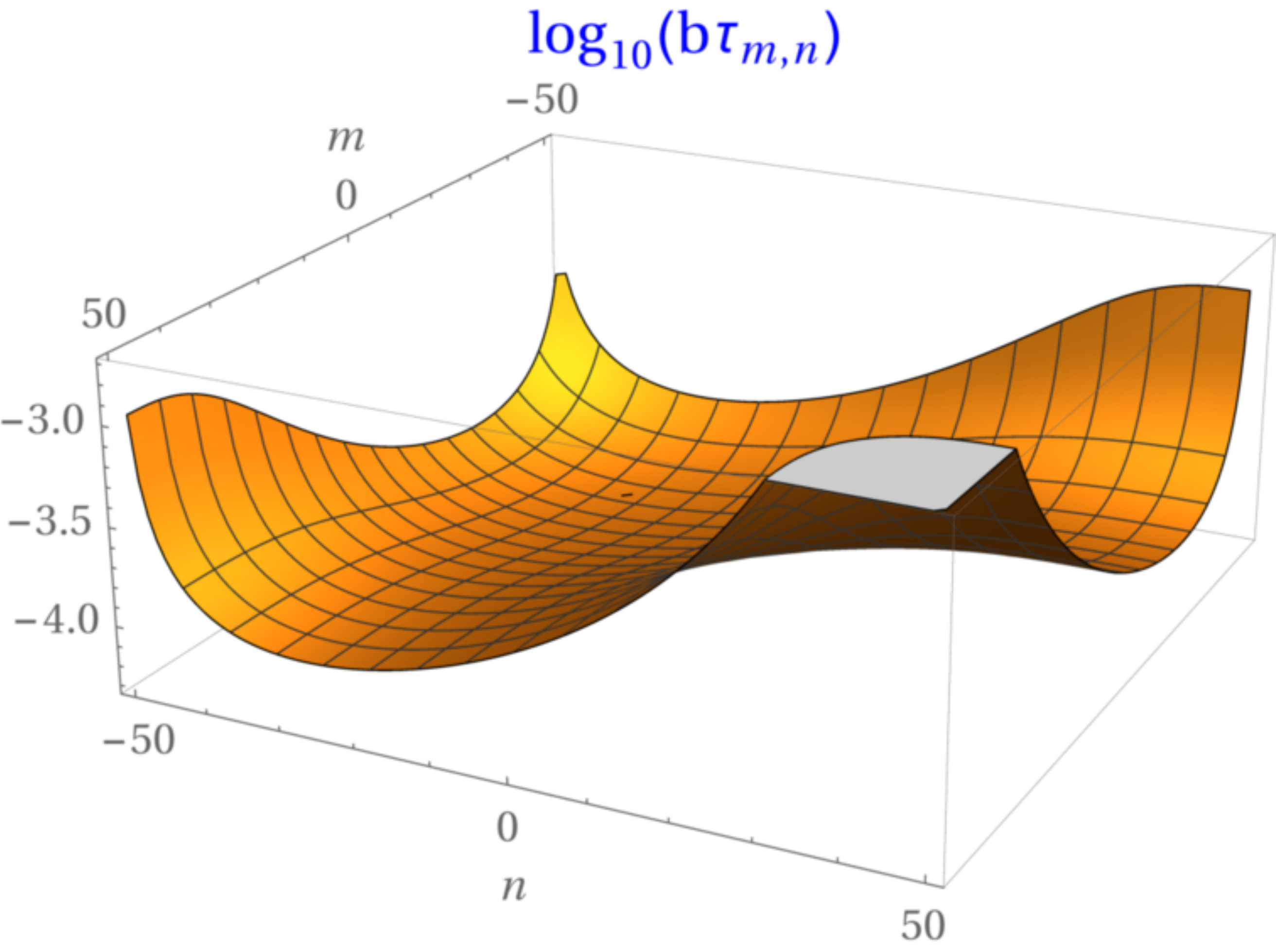}}
      {\includegraphics[width=0.45\textwidth,clip]{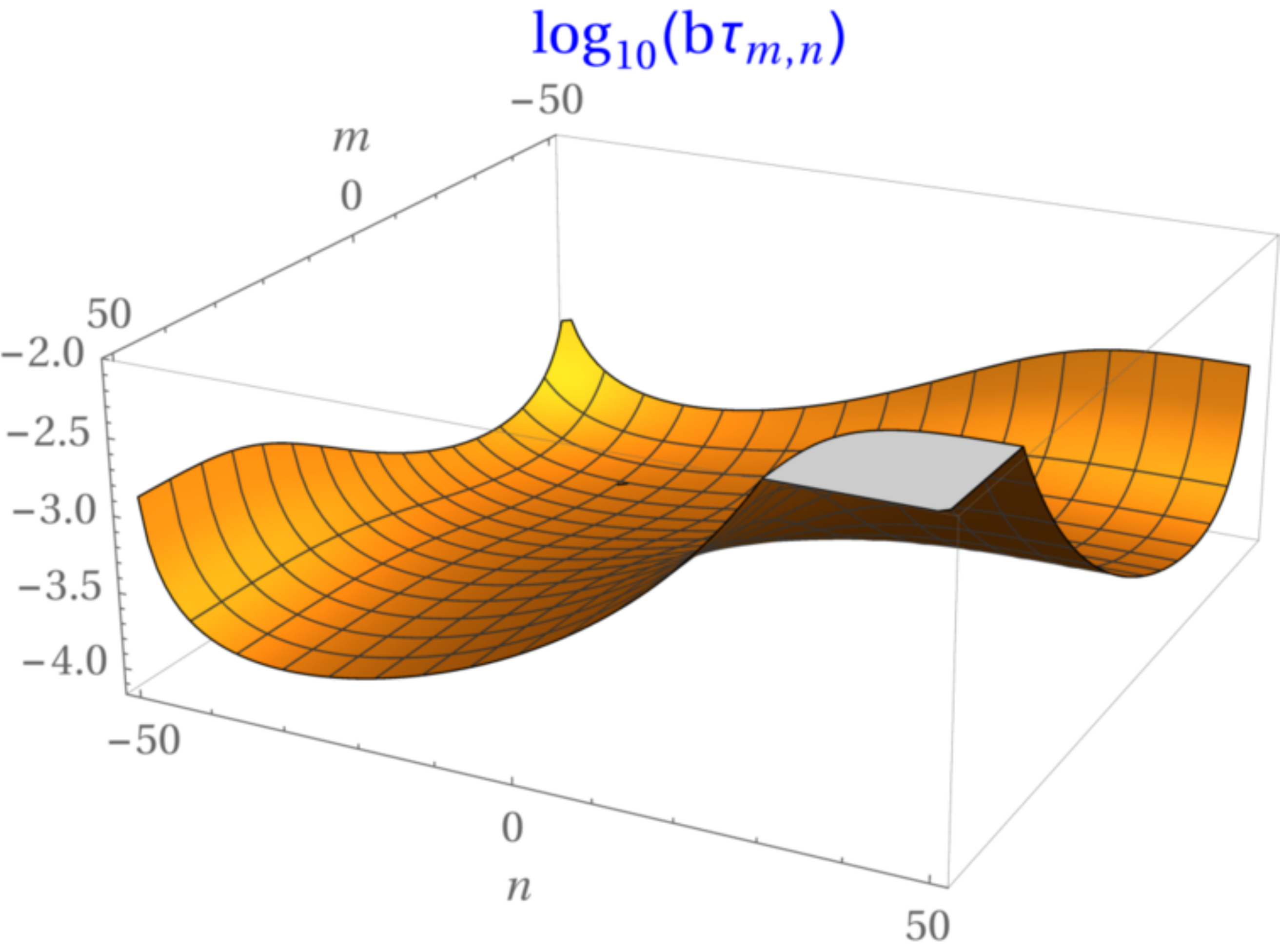}}
      {\includegraphics[width=0.45\textwidth,clip]{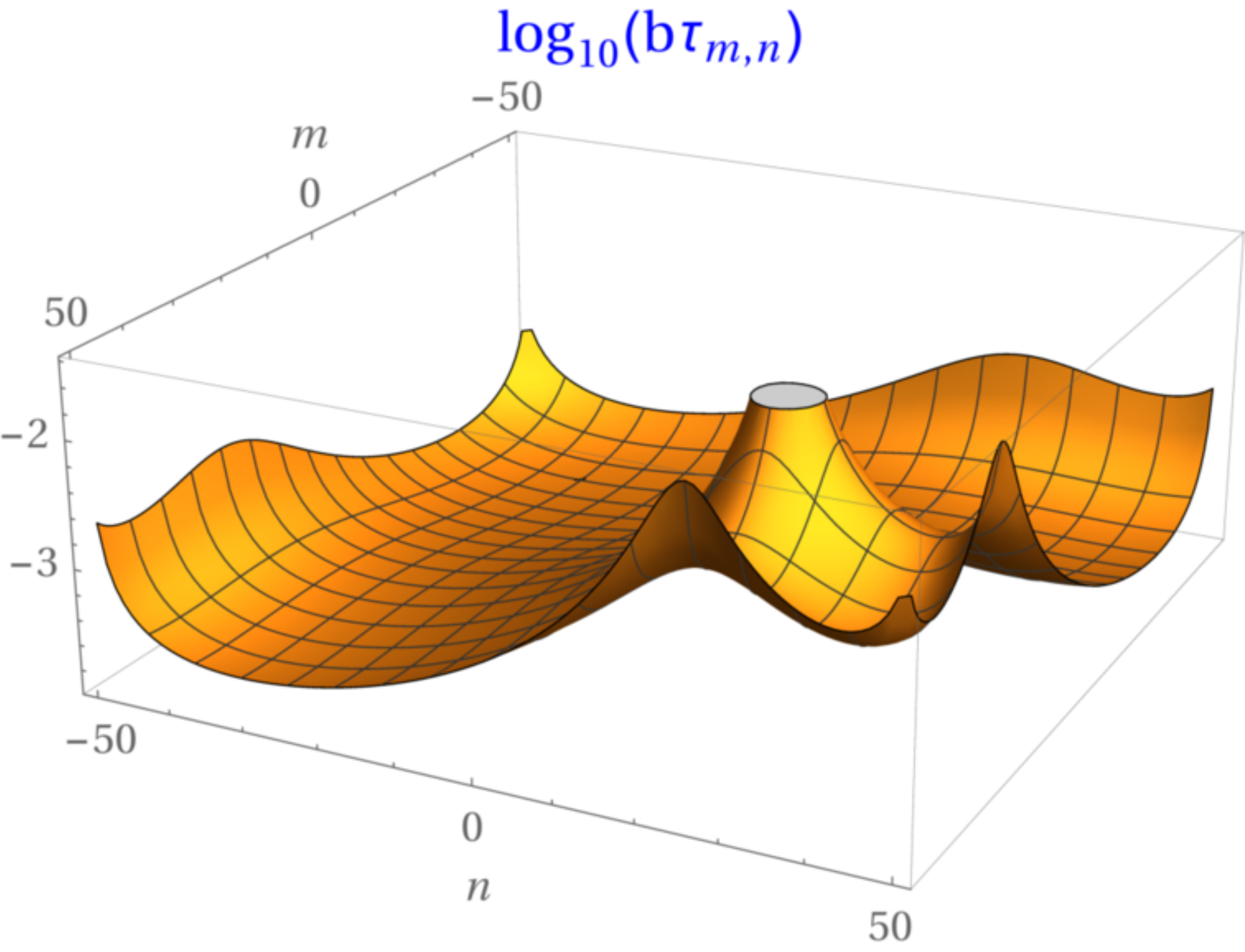}}
      {\includegraphics[width=0.45\textwidth,clip]{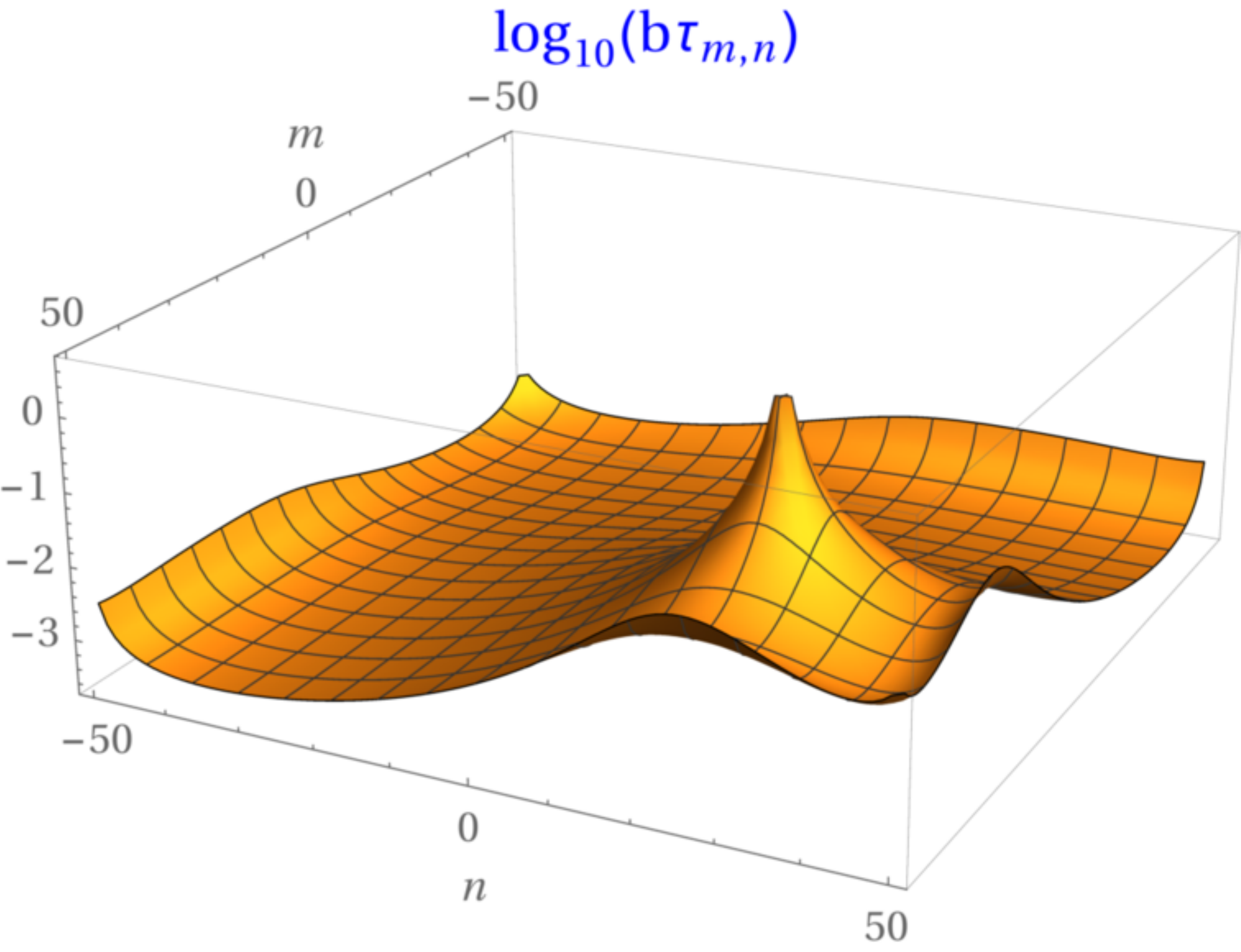}}
      \caption{(Color online) Log plots of the dimensionless
        quantities $b\tau_{m,n}$, where $b=2\gamma\mathcal{J}^3$, as a
        function of $m$ and $n$, with $m\neq n$, obtained from
        Eq.~(\ref{taumnISO}) with $N=100$, $j=N/2$, and, from top to
        bottom, $\Gamma=2\Gamma_c$ (paramagnetic), $\Gamma=\Gamma_c$
        (critical point), and $\Gamma=0.5\Gamma_c$ (ferromagnetic),
        each evaluated at the dimensionless inverse temperatures
        $\beta\mathcal{J}=1$ (left) and $\beta\mathcal{J}=10$
        (right). Notice that the left and right panels are different
        in each case.}
      \label{fig.taumn}
    \end{center}
  \end{figure}

  \begin{figure}[htb]
    \begin{center}
      {\includegraphics[width=0.44\textwidth,clip]{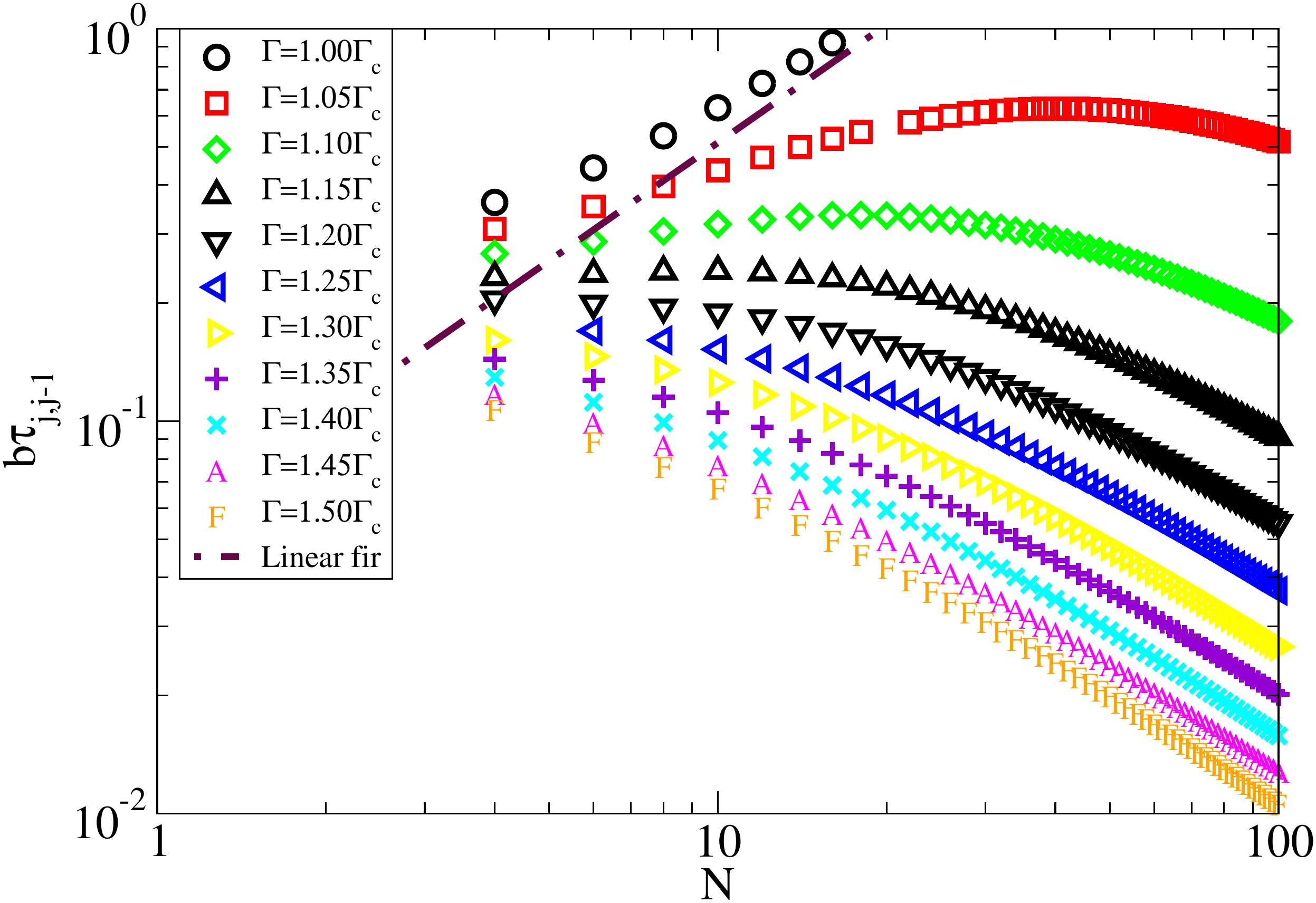}}
      {\includegraphics[width=0.44\textwidth,clip]{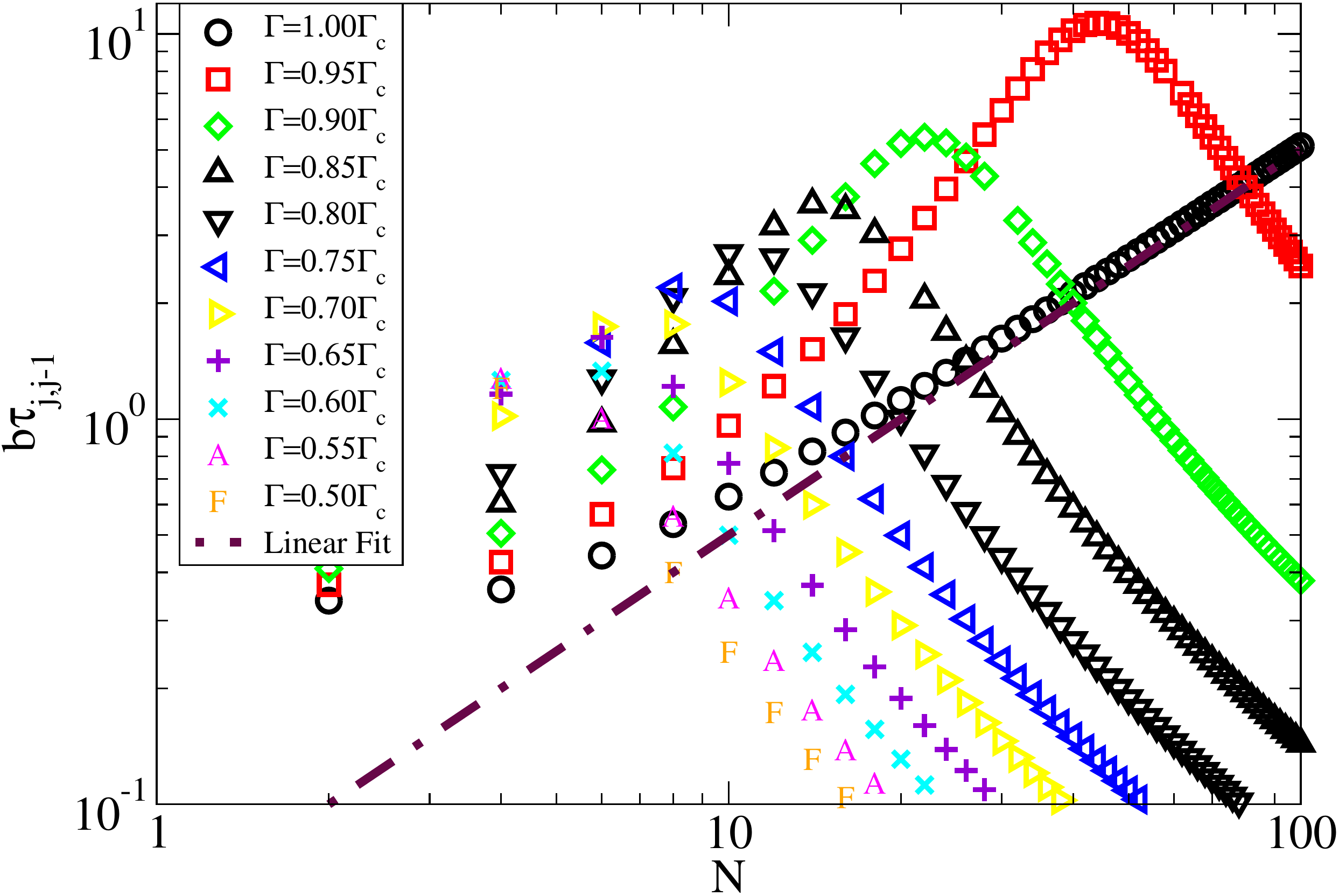}}
      {\includegraphics[width=0.44\textwidth,clip]{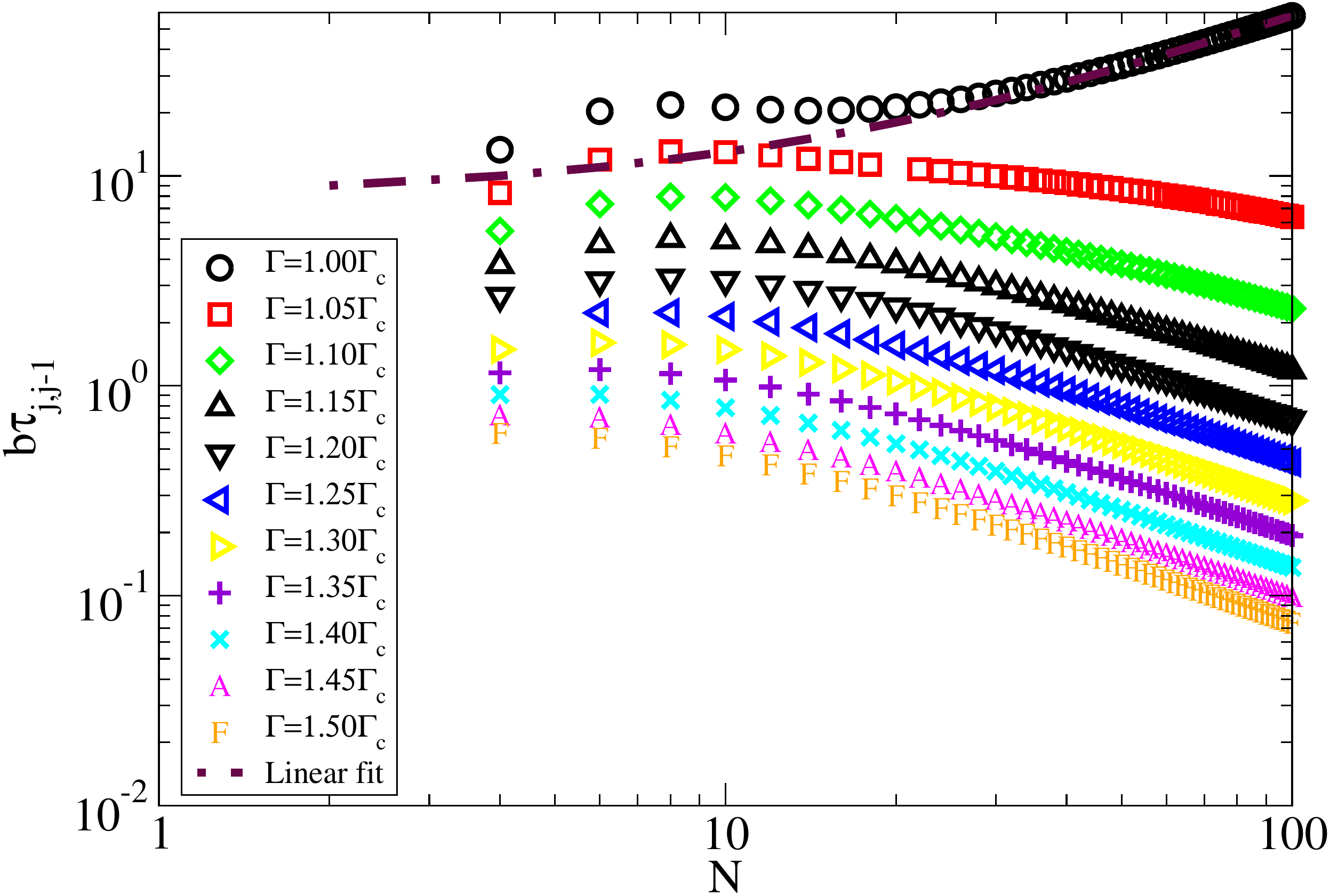}}
      {\includegraphics[width=0.44\textwidth,clip]{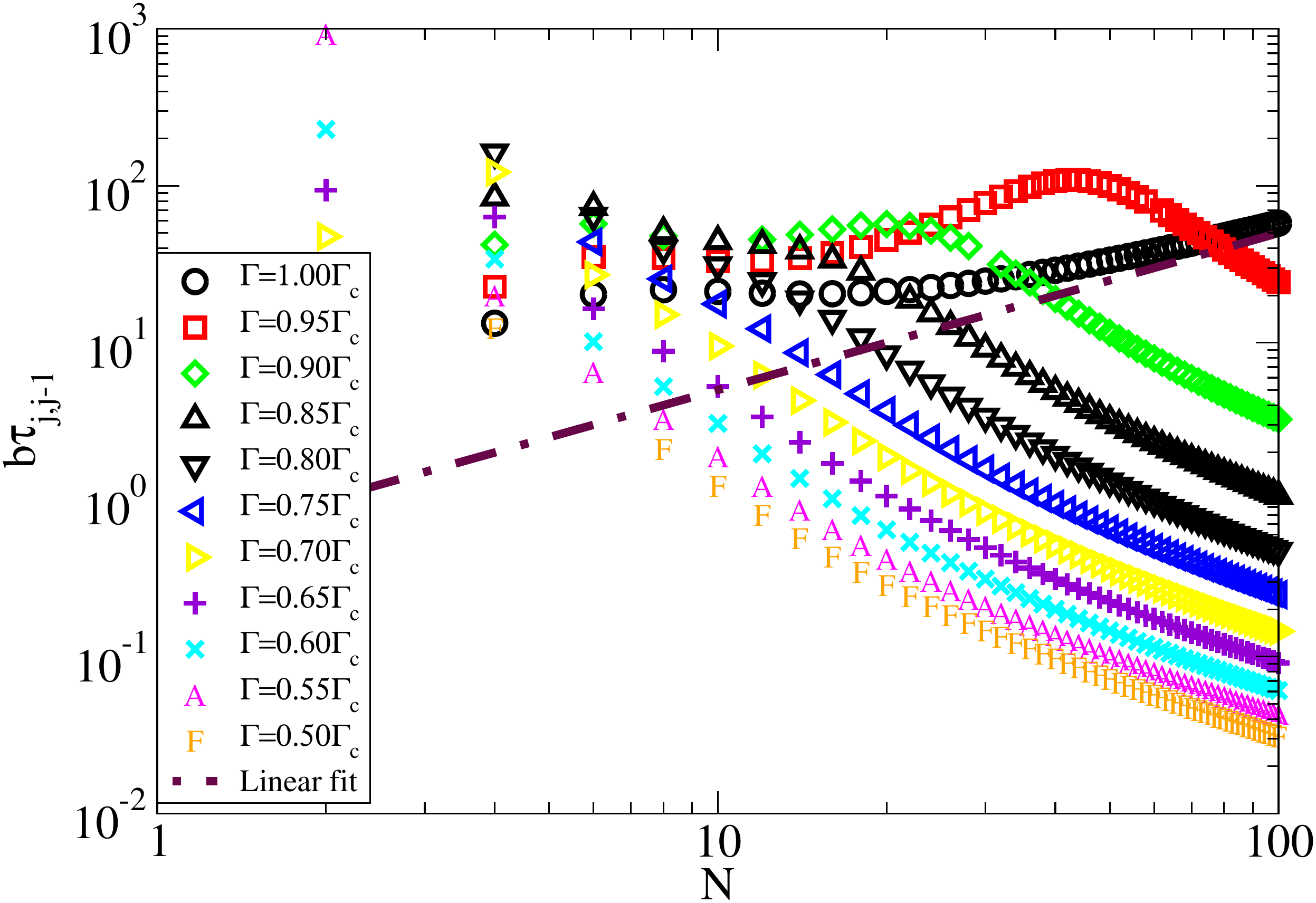}}
      {\includegraphics[width=0.44\textwidth,clip]{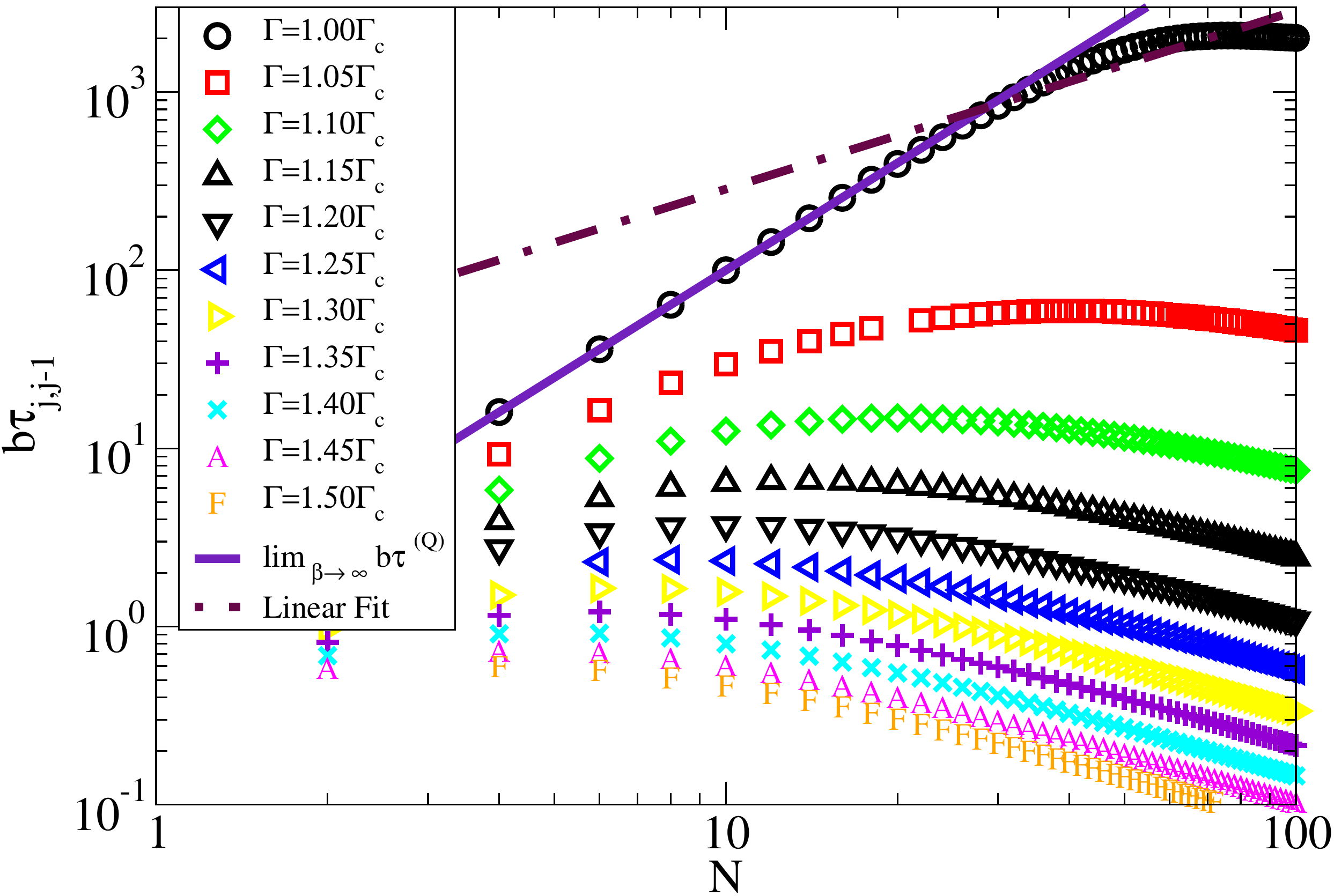}}
      {\includegraphics[width=0.44\textwidth,clip]{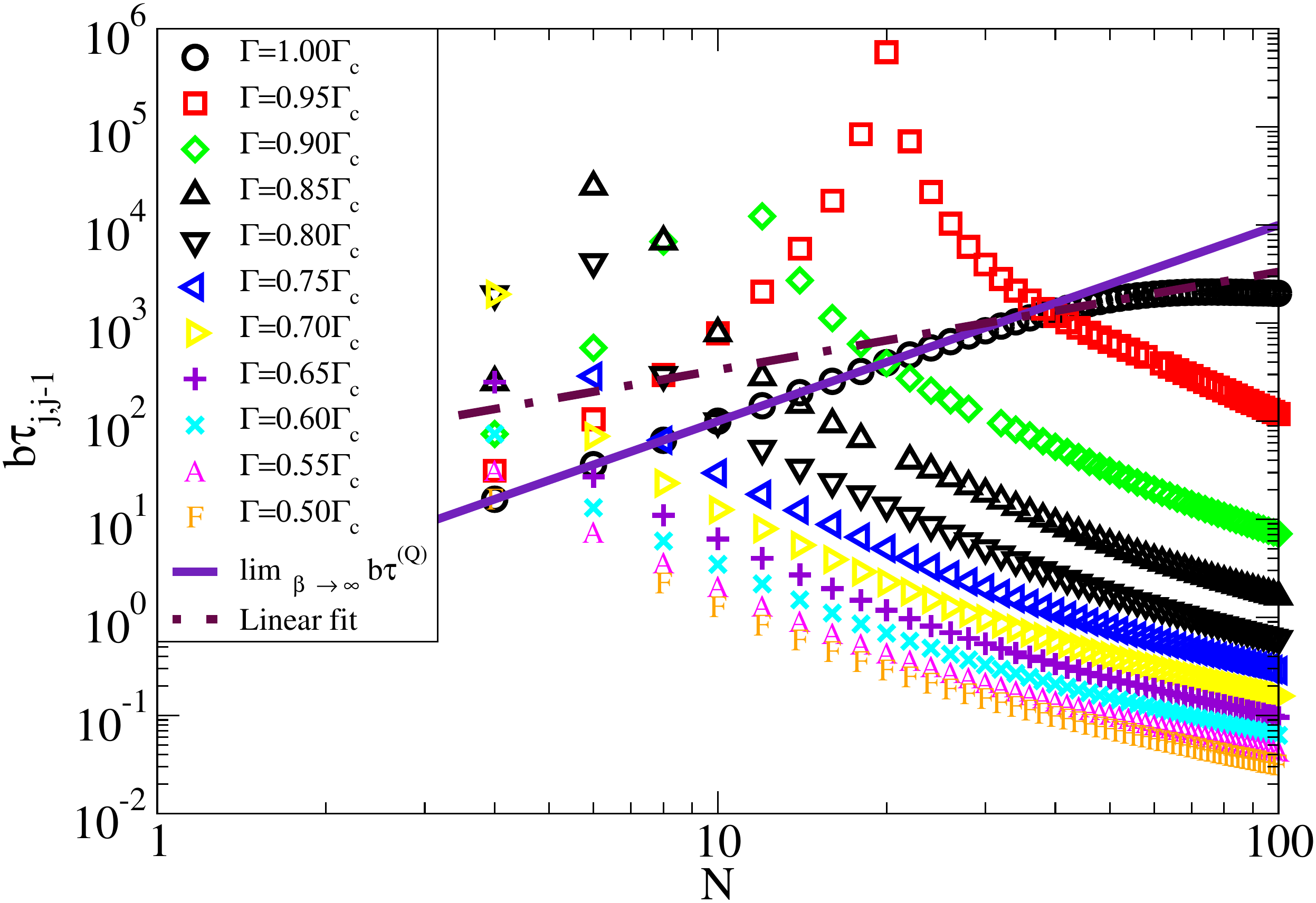}}
      {\includegraphics[width=0.44\textwidth,clip]{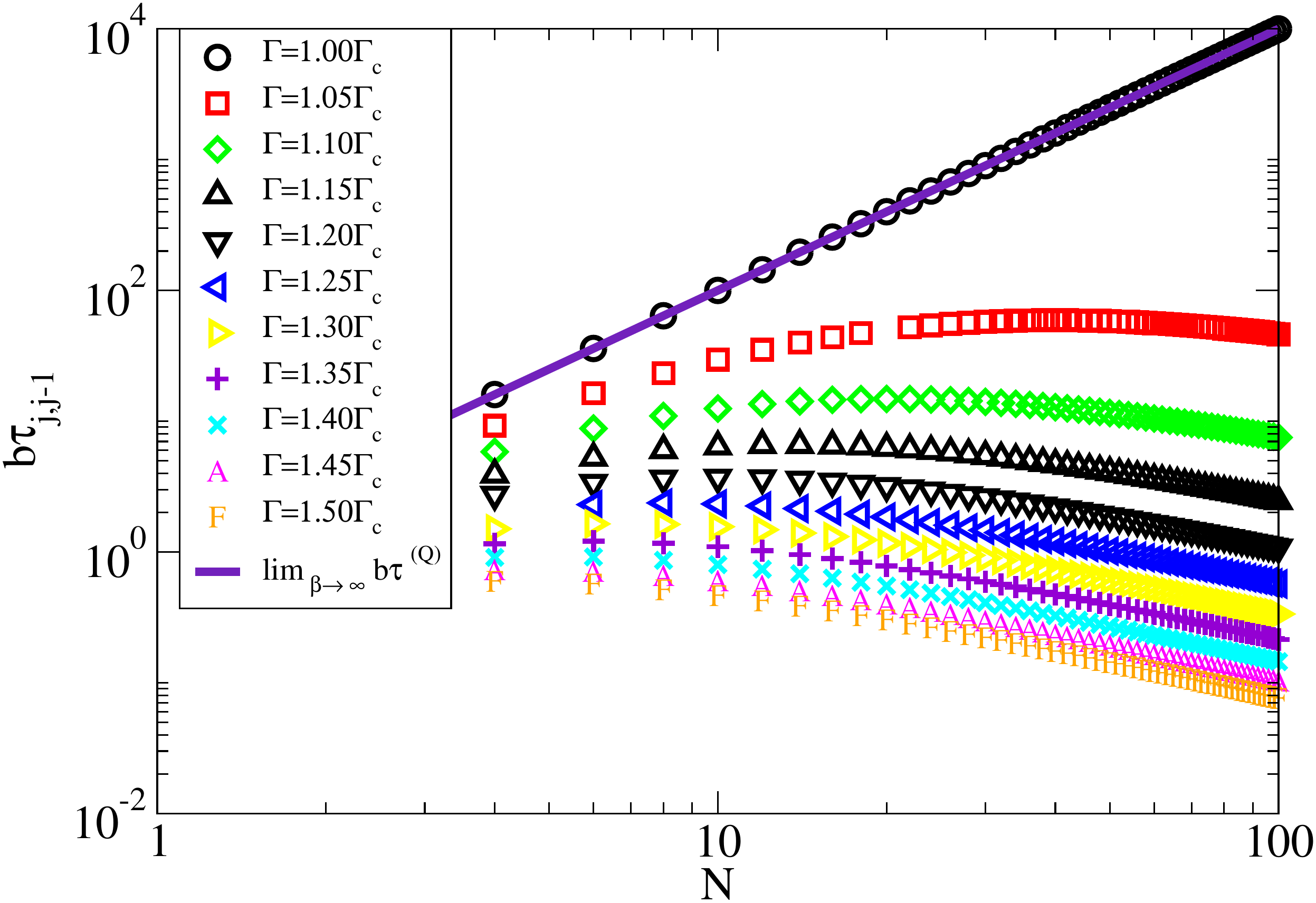}}
      {\includegraphics[width=0.44\textwidth,clip]{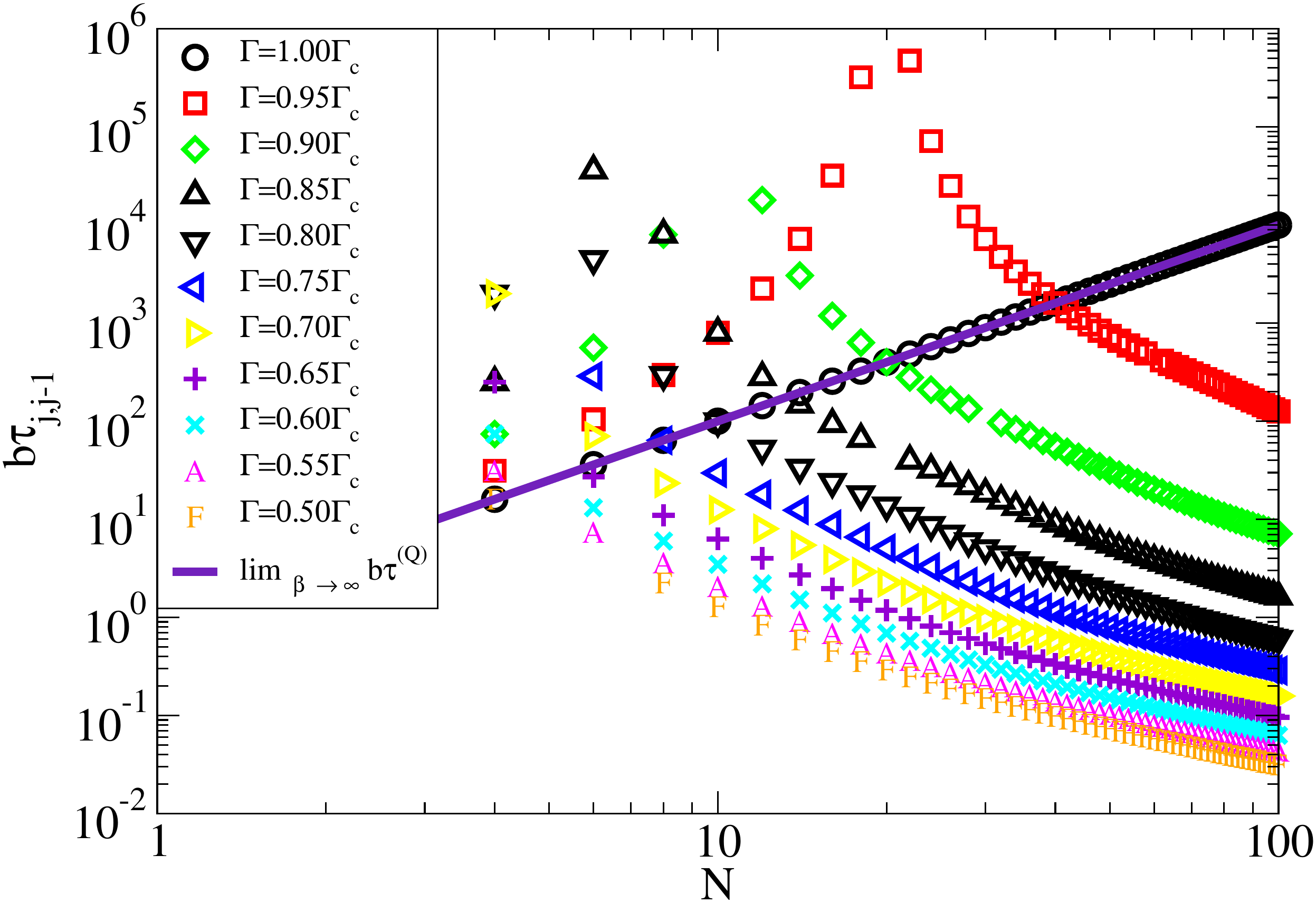}}
      \caption{(Color online) Log-log plots of the dimensionless
        quantities $b\tau_{m=j,n=j-1}$, where
        $b=2\gamma\mathcal{J}^3$, obtained from Eq.~(\ref{taumnISO}),
        as a function of $N$ even, calculated for $j=N/2$, and several
        values of $\Gamma>0$ approaching $\Gamma_c>0$,
        Eq.~(\ref{Gammac}), from above, i.e., in the paramagnetic
        region (left panels), and from below, i.e., in the
        ferromagnetic region (right panels). Different dimensionless
        inverse temperatures are considered, from top to bottom:
        $\beta\mathcal{J}=1$, $\beta\mathcal{J}=10$,
        $\beta\mathcal{J}=100$, and $\beta\mathcal{J}=1000$.  The
        function $\lim_{\beta\to\infty}\tau^{(Q)}$ is obtained from
        Eq.~(\ref{tauQcrit}). Notice however that, by definition,
        $\tau^{(Q)}=\max_{m\neq n}\tau_{m,n}\geq \tau_{j,j-1}$
        (compare Fig. \ref{fig.taumn}).}
      \label{fig.loglog}
    \end{center}
  \end{figure}

  \begin{figure}[htb]
    \begin{center}
      {\includegraphics[width=0.45\textwidth,clip]{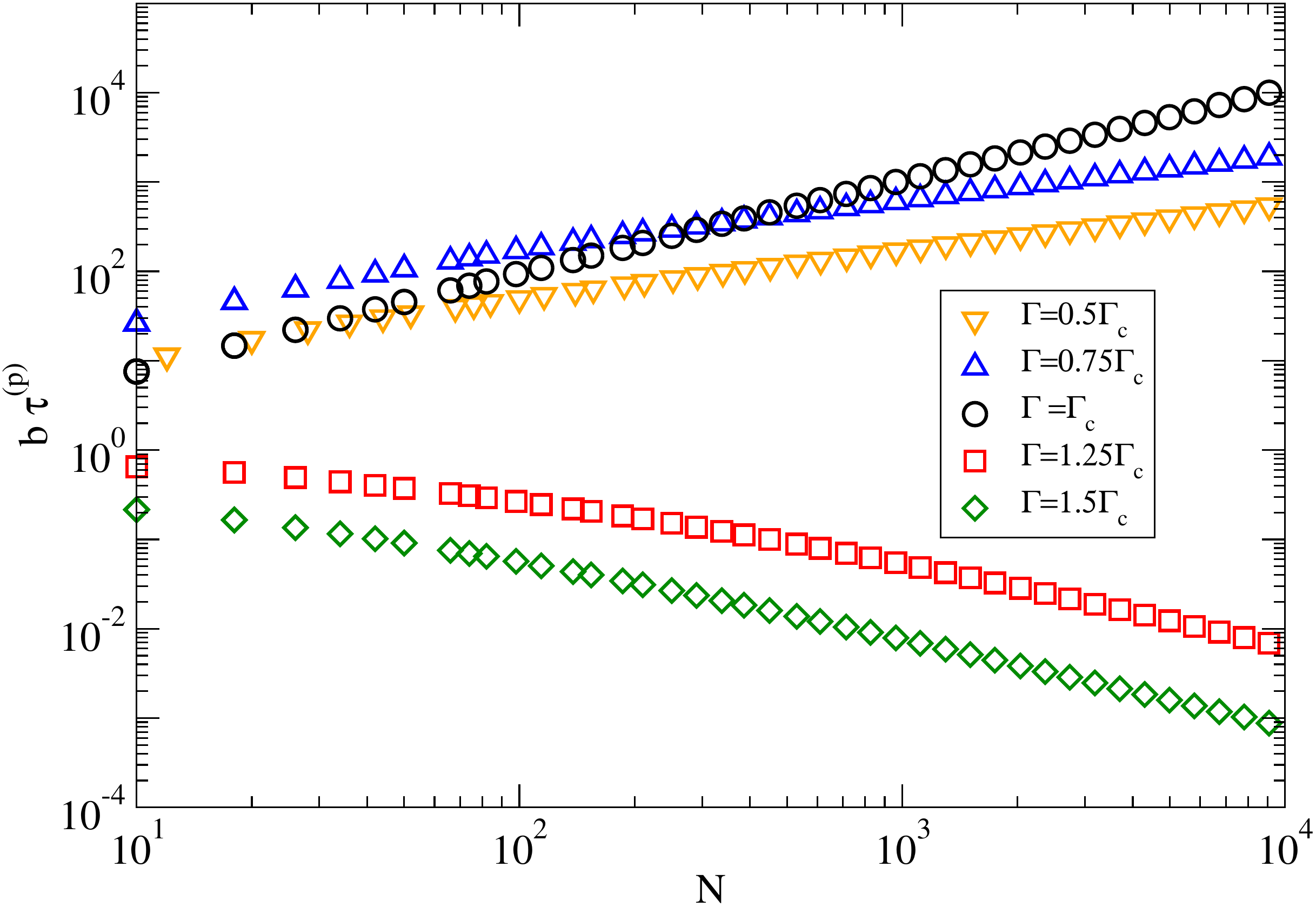}}
      {\includegraphics[width=0.45\textwidth,clip]{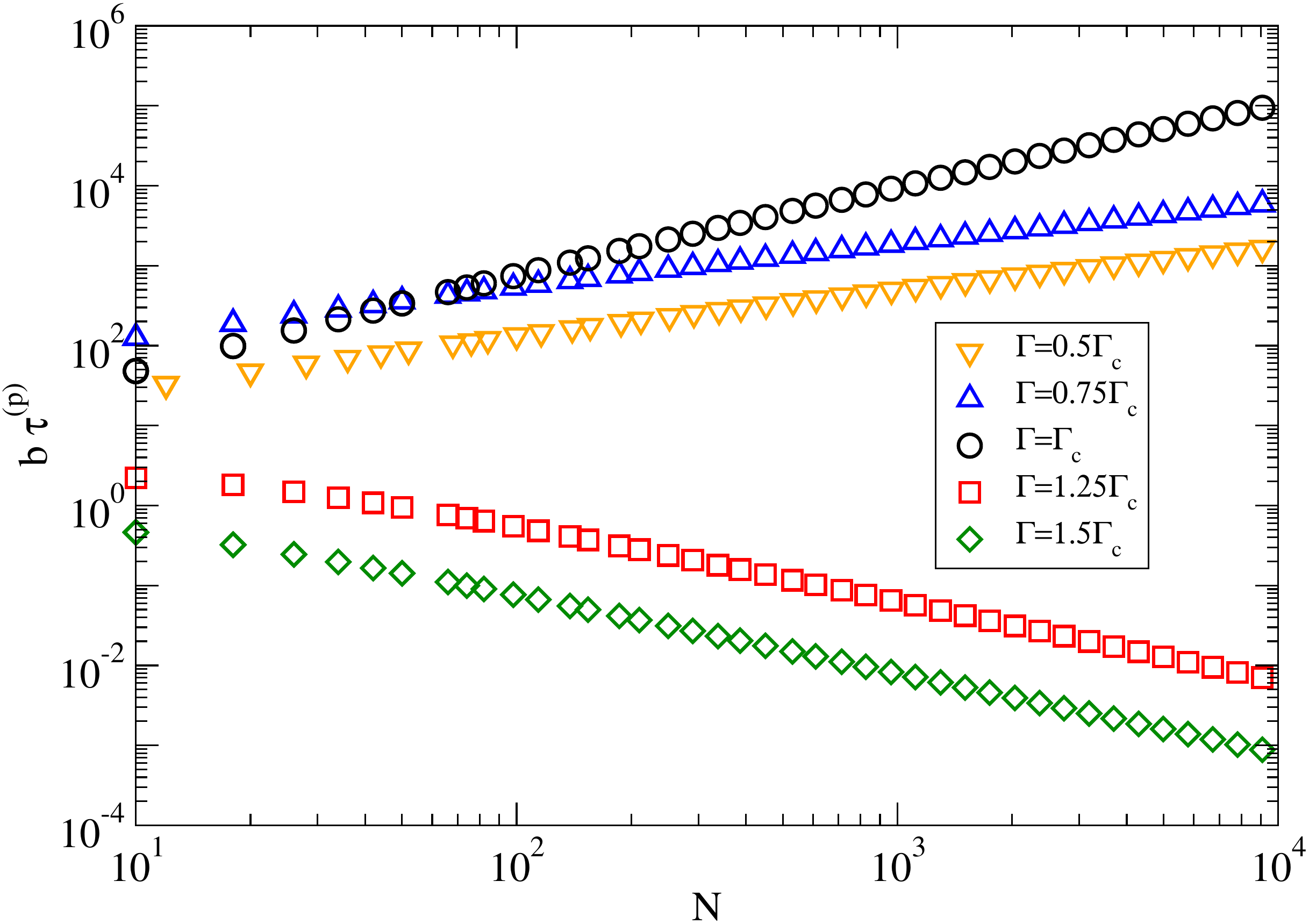}}
      {\includegraphics[width=0.45\textwidth,clip]{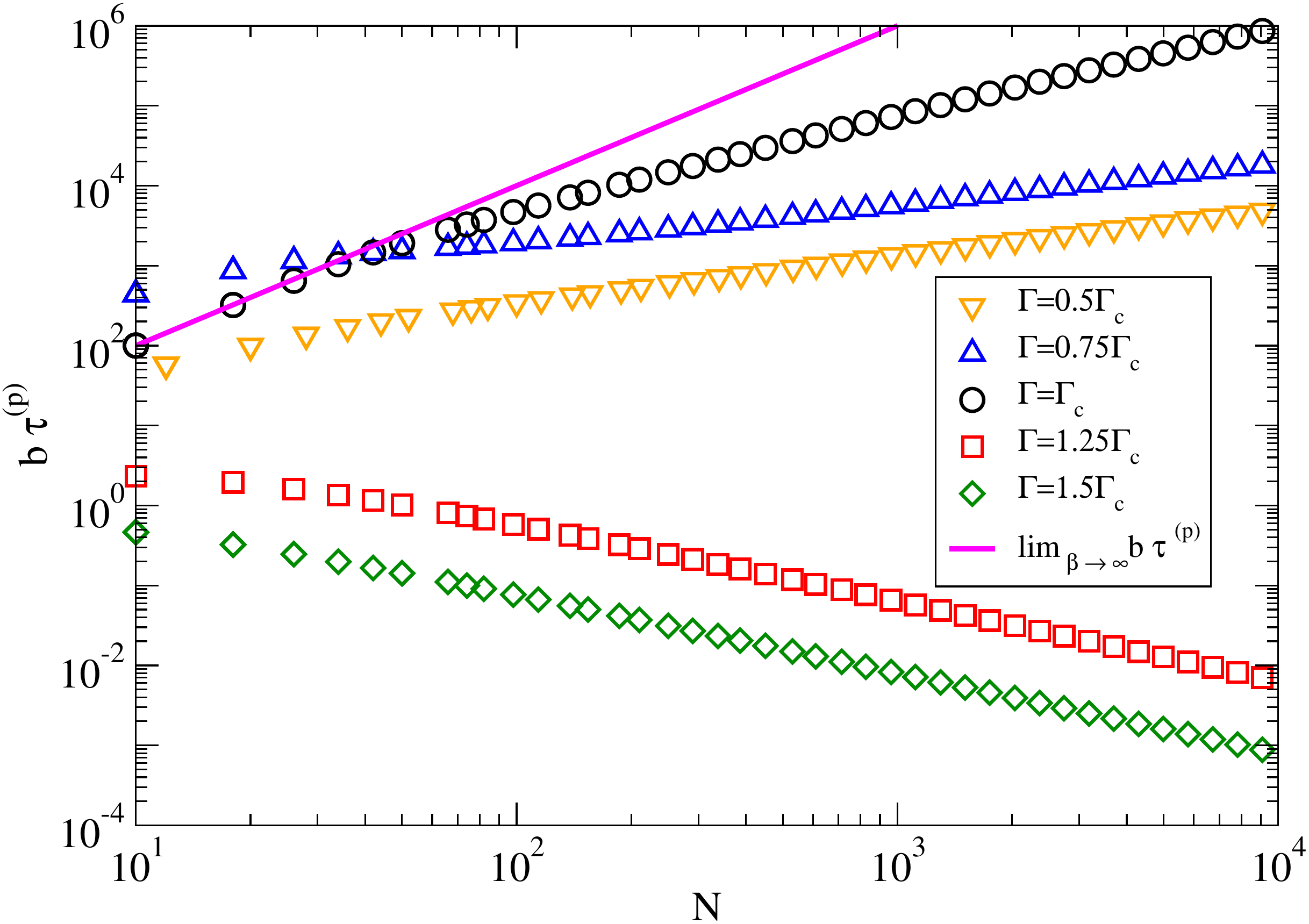}}
      {\includegraphics[width=0.45\textwidth,clip]{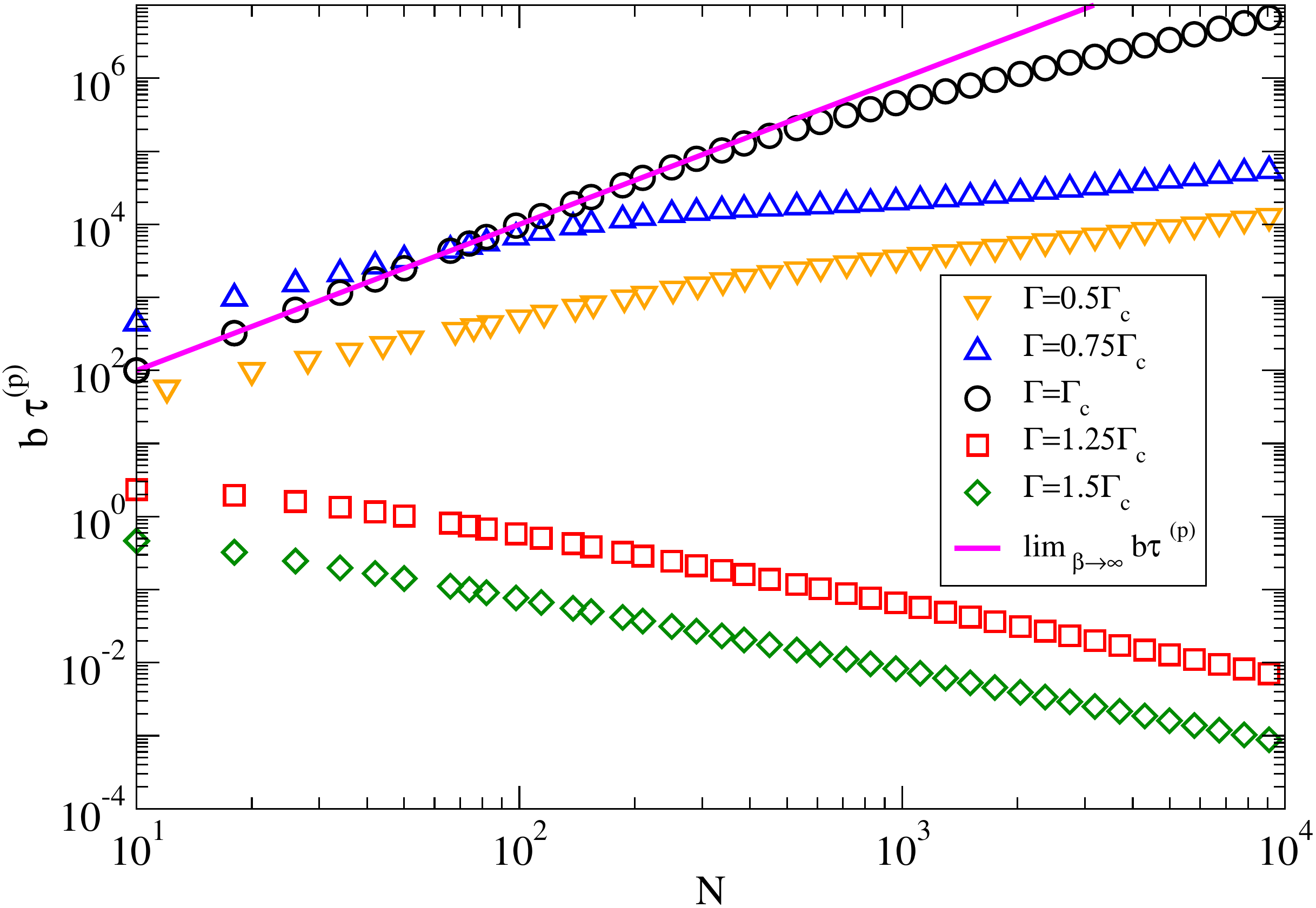}}
      \caption{(Color online) Log-log plots of the dimensionless
        quantity $b\tau^{(P)}=b/\mu_2(\bm{A})$, where
        $b=2\gamma\mathcal{J}^3$ and $\mu_2(\bm{A})$ is the smallest
        non zero eigenvalue of $\bm{A}$, the matrix given by
        Eqs.~(\ref{BISO})-(\ref{AISO}), as a function of $N$ even,
        evaluated for $j=N/2$ and several values of $\beta$ and
        $\Gamma$, above and below the critical point $\Gamma_c$.  Top
        left panel $\beta\mathcal{J}=1$; top right panel
        $\beta\mathcal{J}=10$; bottom left panel
        $\beta\mathcal{J}=100$; bottom right panel
        $\beta\mathcal{J}=1000$.  The function
        $\lim_{\beta\to\infty}\tau^{(P)}$ is given by
        Eq.~(\ref{tauPcrit}) evaluated at $0<\Gamma\leq\Gamma_c$ (it
        provides the same limit in all the ferromagnetic region).  For
        the present values of $\beta$,
        $\lim_{\beta\to\infty}\tau^{(P)}$ matches well with the data
        corresponding to $\Gamma=\Gamma_c$ and $\beta\mathcal{J}=1000$
        when $N\leq 10^3$. For larger values of $\beta$ the agreement
        extends to greater values of $N$ and also to data obtained for
        $\Gamma<\Gamma_c$. }
      \label{fig.tauP}
    \end{center}
  \end{figure}

\end{widetext}

\section{Numerical analysis of isotropic LMG models}
\label{numerical}
We made an exact numerical analysis of Eq.~(\ref{taumnISO}) and of the
eigenvalues of the matrix $\bm{A}$ provided by
Eqs.~(\ref{BISO})-(\ref{AISO}).  The numerical analysis confirms our
analytical formulas and, besides, makes evident the existence of
finite size effects, which are a fingerprint of the phase transition.

Figure~\ref{fig.taumn} provides 3D plots of $\tau_{m,n}$, as a
function of $m$ and $n$, calculated for a few choices of $\Gamma$ and
$\beta$. In agreement with Eqs.~(\ref{taumnISO1}), the maximum of
$\tau_{m,n}$ occurs in correspondence of $m\simeq n \simeq j/2$.

Figure~\ref{fig.loglog} shows the behavior of $\tau_{m=j,n=j-1}$
(i.e., one of the components of the decoherence times $\tau_{m,n}$
close to $\tau^{(Q)}=\max_{m\neq n}\tau_{m,n}$) as a function of the
system size $N$ at different temperatures and for several values of
$\Gamma$ approaching the critical point $\Gamma_c$ in both the
paramagnetic and ferromagnetic regions.  These plots confirm, in
particular, that, for $\beta$ finite, $\tau_{m=j,n=j-1}$ diverges only
at the critical point. More precisely, the divergence is linear in $N$
for $\beta$ sufficiently small, i.e., for $\beta\sim O(1/\Gamma)\sim
O(1/\mathcal{J})$, in agreement with Eq.~(\ref{tauQISO3}). A different
situation occurs instead for $\beta\to\infty$, where the divergence is
quadratic in $N$ and takes place for any $\Gamma$ in the ferromagnetic
region, in agreement with Eq.~(\ref{tauQcrit}).  Fig.~\ref{fig.loglog}
also provides a clear evidence of finite size effects in proximity of
the critical point, which are particularly important in the
ferromagnetic region and at low temperatures.  At some threshold
$N_s(\beta,\Gamma)$, these finite size effects decay approximately as
power laws in $N$ (notice that Fig.~\ref{fig.loglog} is in log-log
scale).  In general, $N_s(\beta,\Gamma)$ turns out to be a non growing
function of $\beta$, whereas, for a given $\beta$, it grows for
$\Gamma$ approaching $\Gamma_c$.

Figure~\ref{fig.tauP} shows $\tau^{(P)}$ as a function of the system
size $N$.  Unlike $\tau^{(Q)}$, we see that, whereas in the
paramagnetic region, $\Gamma>\Gamma_c$, $\tau^{(P)}$ decays as a power
law, in the ferromagnetic region, $\Gamma<\Gamma_c$, $\tau^{(P)}$
grows approximately as a power law even for $\beta$ finite.  Actually,
the behavior of $\tau^{(P)}$ in the ferromagnetic region is not as
smooth as shown in Fig.~\ref{fig.tauP}: by varying $N$ we have
periodic oscillations among three smooth curves associated to
different sequences of $N$ even.  The data shown in
Fig.~\ref{fig.tauP} correspond to one of these sequences, for the
other ones we have a power law growth with a similar exponent but with
a different prefactor.

Another fingerprint of the phase transition that takes place in the
LMG model can be seen in Fig.~\ref{fig.tauP} observing the agreement
between Eq.~(\ref{tauPcrit}) and $\tau^{(P)}$ when the latter is
evaluated at larger and larger values of $\beta$ (see bottom panels).
More precisely, it turns out that, for $N$ sufficiently large,
$\tau^{(P)}(\Gamma_c)\geq \tau^{(P)}(\Gamma)$ for any $\Gamma$, and
that, for $\beta$ sufficiently small, i.e., $\beta \sim
O(1/\Gamma)\sim O(1/\mathcal{J})$, $\tau^{(P)}$ grows no more than
linearly with $N$, while it grows no more than quadratically in $N$
for $\beta$ large.

\section{Conclusions}
\label{conclusions}
We have addressed the thermalization of the LMG model in contact with
a blackbody radiation.  The analysis is done within LBA, a general
mathematical set up developed in~\cite{OP} which allows us to analyze
the thermalization processes of extensive many-body systems.  When
applied to the LMG model embedded in a blackbody radiation, the LBA
equations (which, in the fully coherent regime, coincide with the
QOME) are relatively simple and can be studied analytically in great
detail.  A series of novel results emerge.

First, by analyzing the involved dipole-matrix elements, we find that,
according to the conditions~(\ref{coherent}) and (\ref{incoherent}),
in the general LMG model, i.e., independently of the anisotropy
parameter $\gamma_y$, a full thermalization can take place only if the
density is sufficiently high, while, in the limit of low density, the
system thermalizes partially, namely, within the Hilbert subspaces
$\mathcal{H}_j$ where the total spin has a fixed value.

Second, in the fully coherent regime, and for the isotropic case
$\gamma_y=1$, we are able to perform a comprehensive analysis of the
thermalization. We evaluate the characteristic thermalization time
$\tau$ almost analytically, as a function of the Hamiltonian
parameters and of the system size $N$.

Third, we show that, as a function of $N$, $\tau$ diverges only at the
critical point and in the ferromagnetic region. This divergence is no
more than linear in $N$ for $\beta$ small, and no more than quadratic
in $N$ for $\beta$ large.  In particular, in the ferromagnetic region
and at zero temperature, we prove that $\tau$ diverges just
quadratically with $N$, while quantum adiabatic algorithms lead to an
adiabatic time that diverges with the cube of $N$.

The latter result sheds new light on the problem of preparing a
quantum system in a target state.  If the target state is the GS of a
subspace of the Hilbert space of the system, cooling the system at
sufficiently small temperatures and ensuring, at the same time, that
the system remains sufficiently confined in the chosen subspace, may
produce an arbitrarily accurate result.  This procedure, at least for
the present LMG model coupled to a blackbody radiation, outperforms
the procedure suggested by quantum adiabatic algorithms, where
counterproductive costly efforts are spent to avoid dissipative
effects.  For more general many-body systems, it could be appropriate
to consider cooling processes induced by different, possibly
engineered, thermal reservoirs.  The no-go theorem for exact
ground-state cooling~\cite{nogotheorem}, which apparently prohibits
the application of this idea, can be effectively evaded as discussed
in~\cite{Viola}.

\end{document}